\pdfoutput=1
\documentclass[12pt,preprint]{aastex}

\newcommand{\Rmnum}[1]{\expandafter\@slowromancap\romannumeral #1@}

\shorttitle{Coronal holes, decaying active regions and global coronal structure}
\shortauthors{Petrie and Haislmaier}

\begin{document}


\title{Low-latitude coronal holes, decaying active regions and global coronal magnetic structure}


\author{G.J.D. Petrie$^1$ \& K.J. Haislmaier$^2$}
\affil{$^1$National Solar Observatory, Tucson, AZ 85719, USA\\
$^2$George Mason University, Fairfax, VA 22030, USA}



\begin{abstract}
We study the relationship between decaying active region magnetic fields, coronal holes and the global coronal magnetic structure using Global Oscillations Network Group (GONG) synoptic magnetograms, Solar Terrestrial RElations Observatory (STEREO) extreme ultra-violet (EUV) synoptic maps and coronal potential-field source-surface (PFSS) models. We analyze 14 decaying regions and associated coronal holes occurring between early 2007 and late 2010, four from cycle 23 and 10 from cycle 24. We investigate the relationship between asymmetries in active regions' positive and negative magnetic intensities, asymmetric magnetic decay rates, flux imbalances, global field structure and coronal hole formation. Whereas new emerging active regions caused changes in the large-scale coronal field, the coronal fields of the 14 decaying active regions only opened under the condition that the global coronal structure remained almost unchanged. This was because the dominant slowly-varying, low-order multipoles prevented opposing-polarity fields from opening and the remnant active-region flux preserved the regions' low-order multipole moments long after the regions had decayed. Thus the polarity of each coronal hole necessarily matched the polar field on the side of the streamer belt where the corresponding active region decayed. For magnetically isolated active regions initially located within the streamer belt, the more intense polarity generally survived to form the hole. For non-isolated regions, flux imbalance and topological asymmetry prompted the opposite to occur in some cases.
\end{abstract}


\section{Introduction}
\label{sect:intro}
\indent 

The solar coronal magnetic field is formed by active regions emerging from the solar interior and becoming distributed throughout the atmosphere by flux transport processes (e.g., Mackay and Yeates~2012). The field can become open to the heliosphere in regions where a single magnetic polarity dominates, including the polar caps and other regions of isolated polarity (Harvey and Recely~2002). Wherever the coronal fields extend to great heights, the magnetic tension force can be overcome by the thermal pressure of the expanding coronal plasma, and the field dragged out and opened to form a coronal hole (Parker~1958). In extreme ultraviolet (EUV) and soft X-rays (SXR) coronal holes appear as dark regions whereas in helium images they appear bright.

Harvey and Recely~(2002) identified three classes of coronal hole: polar coronal holes confined to high latitudes ($>60^{\circ}$, $<-60^{\circ}$); isolated coronal holes at active latitudes associated with the remnant fields of decayed active regions; and transient coronal holes that briefly form after coronal mass ejections. From a study of the polar coronal holes during cycles 22 and 23, Harvey and Recely concluded that polar coronal holes are formed from isolated holes associated with decayed active-region field of trailing polarity. Earlier, Harvey and Sheeley~(1979) had shown that during the ascending phase of a solar cycle the leading polarities of active regions may diffuse to form locally unbalanced flux patterns that tend to form coronal holes ahead of active regions.

The declining phase of cycle 23 roughly coincided with the beginning of NSO's Global Oscillations Network Group (GONG) synoptic magnetogram program and the launch of NASA's Solar Terrestrial RElations Observatory (STEREO) mission (Kaiser et al.~2008) in late 2006. The cycle 23 minimum produced unusually many coronal holes at active latitudes compared to previous cycle minima, mainly owing to the weakness of the polar fields during cycle 23 (de Toma~2011). The decline of solar cycle 23 and the ascent of cycle 24 have provided excellent conditions for studying isolated coronal holes.

Using Mt. Wilson Observatory and Wilcox Solar Observatory synoptic magnetograms and STEREO EUV synoptic maps, Wang et al.~(2010) studied three instances of small coronal holes appearing at the edges of newly emerged active regions and then expanding and ultimately joining the polar coronal holes. They were able to simulate the formation of three coronal holes using kinematic photospheric flux transport models in combination with extrapolated potential-field source-surface (PFSS) models for the coronal field. They found that the coronal holes could be explained by a combination of photospheric flux transport processes: turbulent diffusion, differential rotation and the interaction between neighboring active regions. They concluded that the global distribution of photospheric flux determined where the coronal holes formed, in particular the proximity or otherwise of strong flux of the opposite polarity. In the PFSS model, coronal loops are assumed to become open if they extend beyond a certain height, hence open field regions form wherever there is insufficient photospheric flux of the opposite polarity to connect to.

Karachik et al.~(2010) used Michelson Doppler Imager (MDI) and GONG full-disk magnetograms and Extreme-ultraviolet Imaging Telescope (EIT) EUV images to investigate four cases of isolated polar coronal holes forming on the remnants of decaying active regions. In contrast to Wang et al.~(2010), Karachik et al.~(2010) focused on the evolution of individual bipolar active regions. They measured several parameters of the active regions over successive rotations including the imbalance and the compactness of the magnetic flux. They found that less compact field tends to decay faster than more compact field, causing a gradual increase in flux imbalance in the region and the formation of a coronal hole at the location of the more compact flux. Karachik et al.'s four cases occurred during the ascent of cycle 24 in 2009 and their conclusions are consistent with those of Sheeley and Harvey~(1979). Using extrapolated PFSS models Karachik et al. found that in three cases, all of them holes of positive polarity, some field rooted at the coronal hole locations connected back to the photosphere close to the north polar coronal hole. These fields therefore appeared to play a role in the solar cycle process, evolving from toroidal active region field into poloidal field.

In this paper we follow Karachik et al.~(2010) by focusing on the relationship between the decaying fields of active regions and the formation of coronal holes. We investigate which factors can determine the formation of the coronal holes including active region flux imbalance, field strength, compactness, interactions with other fields, and the structure of the coronal field represented by PFSS models. We also try to clarify the role of these fields in the solar activity cycle. Most of our examples come from the ascending phase of cycle 24 with a minority of cases representing the declining phase of cycle 23. The paper is organized as follows. We describe the observational data in Section~\ref{sect:data}. The 14 pairs of active regions and coronal holes will be introduced and their properties summarized in Section~\ref{sect:arch}. PFSS models will be used to aid interpretation of coronal hole formation in Section~\ref{sect:pfss}. We will conclude in Section~\ref{sect:conclusion}.

\section{Data}
\label{sect:data}

Full-disk images of the relative Doppler shift of the Ni~{\sc i} line at 676.8~nm are available from each of the six GONG telescopes at a cadence of 1~minute, weather permitting. GONG's six stations together provide round-the-clock coverage with approximately an 87\% duty cycle. The spatial sampling of the GONG images is $2.^{\prime\prime}5$~pixel$^{-1}$ and the instrumental sensitivity is about 3~G~pixel$^{-1}$. Since the GONG instrumentation was upgraded in 2006, resulting in a great reduction of the zero-point error in the field measurements, synoptic maps for the radial component of the photospheric field have been constructed for every Carrington rotation.

The STEREO Sun Earth Connection Coronal and Heliospheric Investigation (SECCHI) instrument package (Howard et al.~2008) has four instruments: an extreme ultraviolet imager, two white-light coronagraphs and a heliospheric imager. The extreme ultraviolet imager, EUVI (Wuelser et al.~2004) observes in four spectral channels that span the 0.1 to 20~MK temperature range. In this paper we use synoptic maps based on 19.5~nm images, corresponding to about 2.5~MK. Coronal holes tend to appear clearest in these maps.

We searched for decaying active regions and associated coronal holes by visually inspecting photospheric synoptic magnetograms from GONG and the corresponding synoptic EUV maps from the STEREO A spacecraft's EUVI instrument. The GONG synoptic map series begins with Carrington rotation (CR) 2047 (August 2006) whereas the STEREO maps begin at CR2051 (early 2007). From pairs of GONG and STEREO synoptic maps we identified 14 cases where the decay of an active region could confidently be associated with the formation of a coronal hole. These include four examples from the declining phase of cycle 23 and ten from the ascent of cycle 24. These are listed in Table~\ref{arlist} and examples are shown in Figures~\ref{fig:ex3} and \ref{fig:ex6}. We only consider examples where active regions could be identified over at least two consecutive rotations and where the remnant flux of the decaying active region could be associated with a newly formed coronal hole in the corresponding STEREO map.

GONG is a ground-based network of telescopes whereas the STEREO A and B spacecraft orbits lead and lag the Earth in its orbit around the Sun. Since their launch in August 2006, the twin STEREO spacecraft have been separating at a rate of $46^{\circ}$ per year. The observations analyzed in this paper date from early 2007 until late 2010, over which time the separation between STEREO A and Earth grew from a fraction of a degree to about 85$^{\circ}$, representing a time delay between a GONG observation of a given longitude and the corresponding STEREO A observation of that same longitude of up to 6.5 days. Because of this time delay the GONG and STEREO maps cannot be treated as being nearly simultaneous, but since we are studying the slow decay of active regions and formation of coronal holes over multiple rotations the pairs of GONG and STEREO synoptic maps are still applicable. Our approach differs from that of Karachik et al.~(2010) in that we do not estimate coronal hole magnetic fluxes by deriving coronal boundaries from STEREO EUV data and mapping them onto GONG magnetograms. Instead we study the magnetic fluxes of the decaying active regions.

\section{Active Regions and Coronal Holes}
\label{sect:arch}

\begin{figure} 
\begin{center}
\resizebox{0.45\textwidth}{!}{\includegraphics*{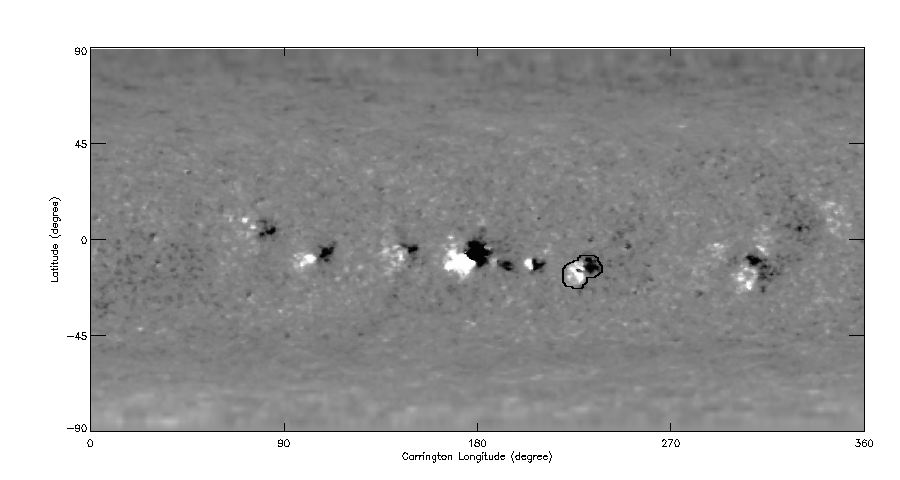}}
\resizebox{0.45\textwidth}{!}{\includegraphics*{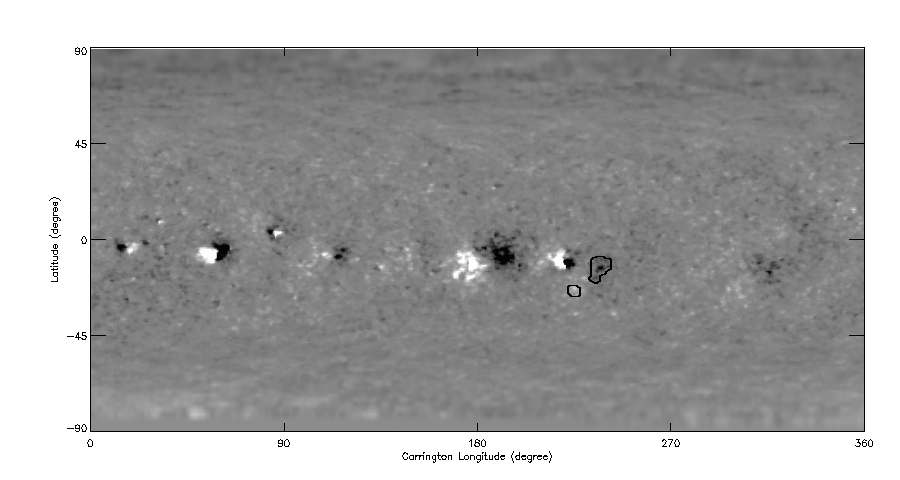}}
\resizebox{0.45\textwidth}{!}{\includegraphics*{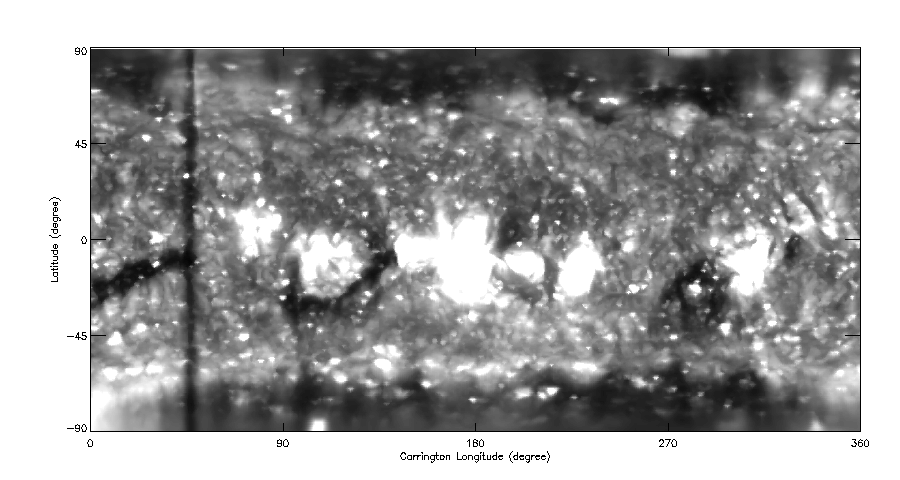}}
\resizebox{0.45\textwidth}{!}{\includegraphics*{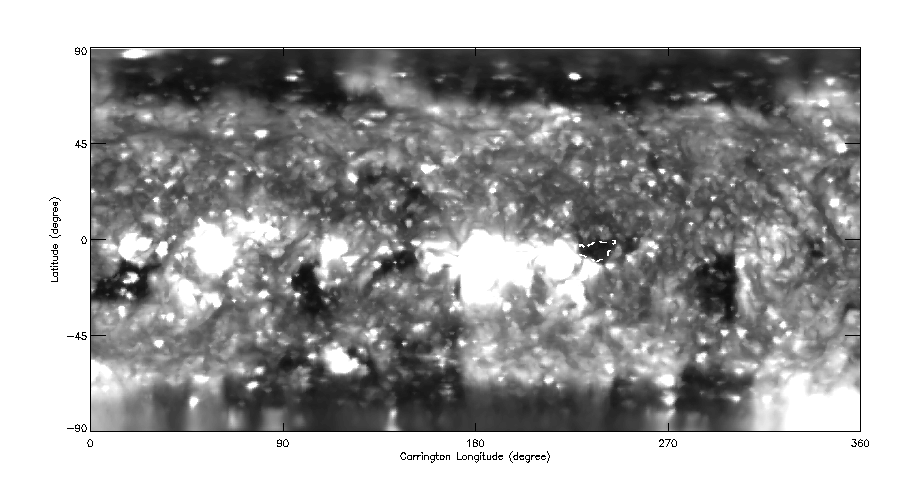}}
\resizebox{0.45\textwidth}{!}{\includegraphics*{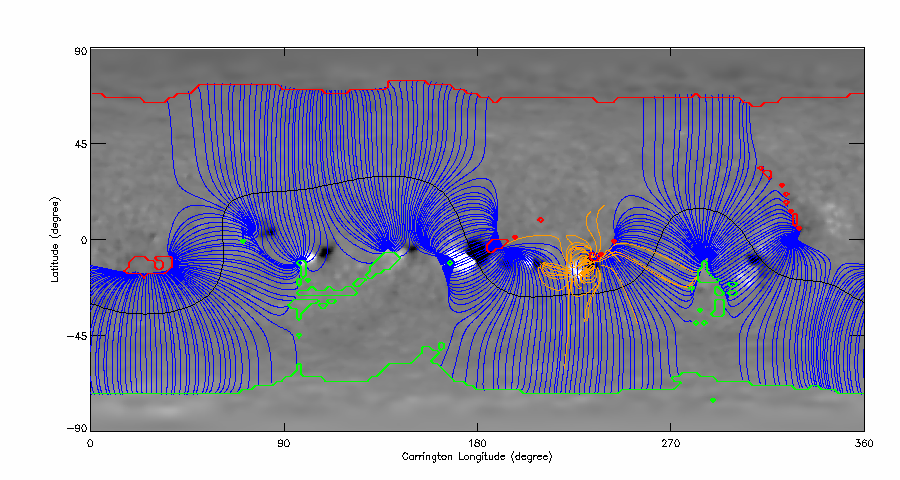}}
\resizebox{0.45\textwidth}{!}{\includegraphics*{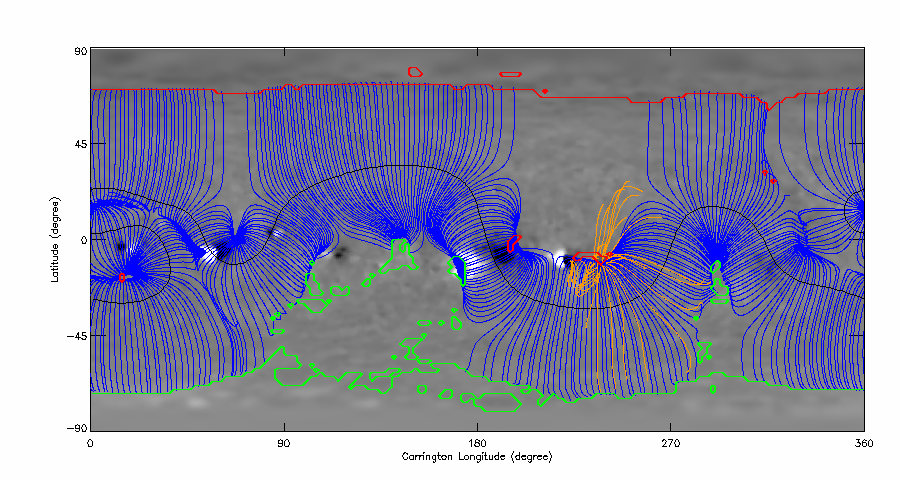}}
\resizebox{0.45\textwidth}{!}{\includegraphics*{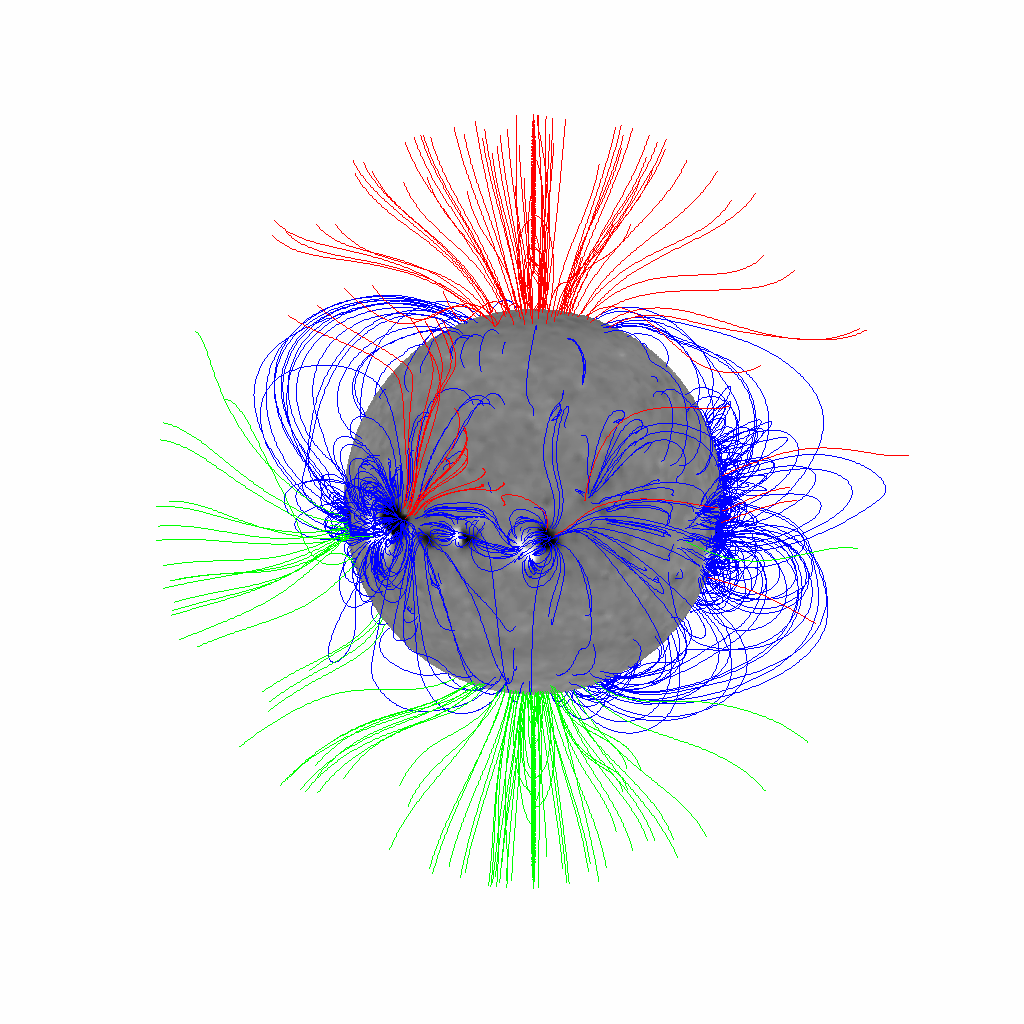}}
\resizebox{0.45\textwidth}{!}{\includegraphics*{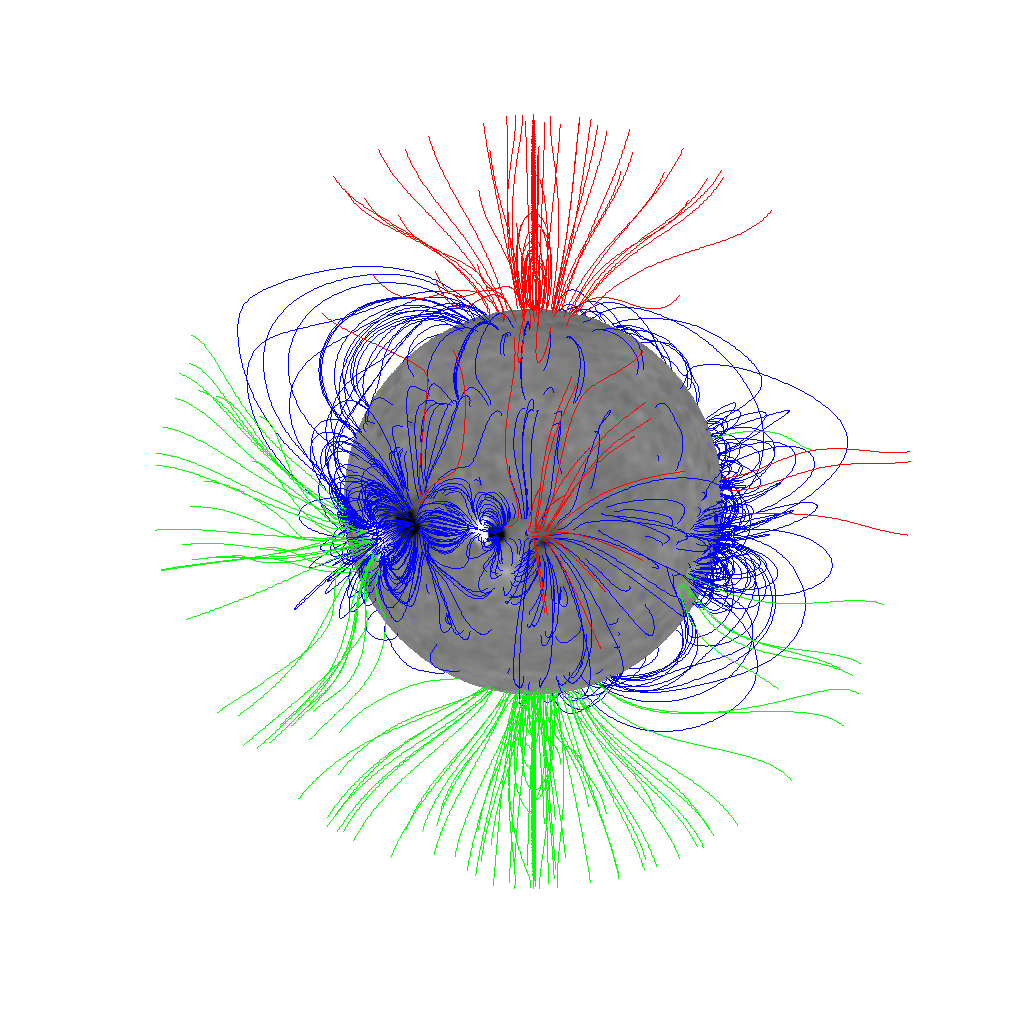}}

\vspace{-1.5cm}
\end{center}
\caption{\footnotesize Example 3 of Table~\ref{arlist}. Top row: GONG synoptic magnetograms for Carrington rotations 2057 and 2058 with positive/negative flux in bright/dark greyscale. The decaying active region is indicated by the black contour curves. Second row: the STEREO EUVI 195~\AA\ synoptic maps for Carrington rotations 2057 and 2058. The coronal hole is indicated by the white dashed EUV brightness contour curve. Third row: PFSS models over-plotted on the synoptic magnetograms. The green and red contours indicate foot-point locations of positive and negative coronal holes. The blue lines represent the streamer belt, and the black source-surface neutral line the equatorial current sheet. The yellow lines represent the active region fields whose foot-points lie within the contour curves shown in the top row of plots. Plots of the same models in spherical coordinates are shown in the bottom row. Closed lines are plotted in blue and positive/negative open lines in green/red. The spherical coordinate system is rotated so that the decaying region is at central meridian.}
\label{fig:ex3}
\end{figure}

\begin{figure} 
\begin{center}
\resizebox{0.45\textwidth}{!}{\includegraphics*{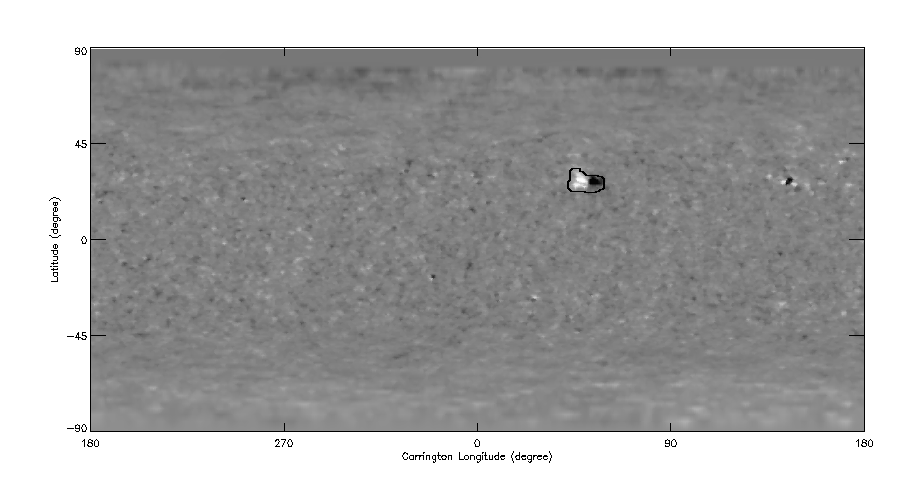}}
\resizebox{0.45\textwidth}{!}{\includegraphics*{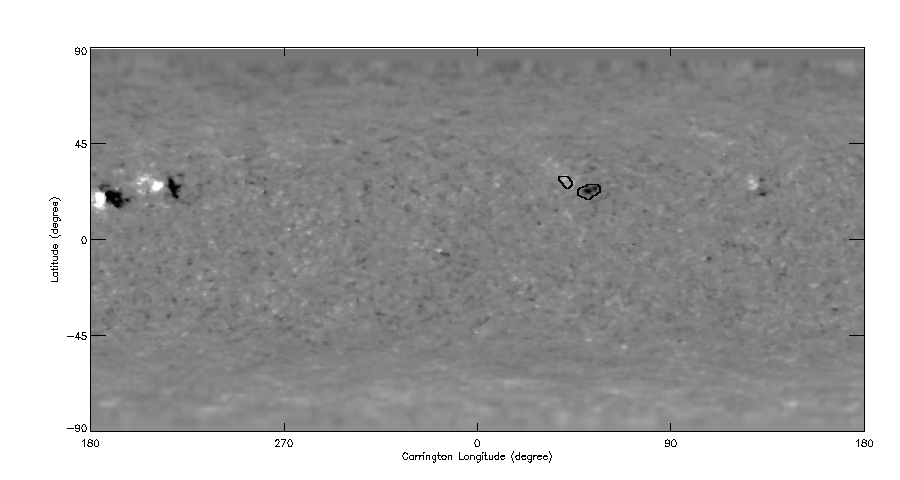}}
\resizebox{0.45\textwidth}{!}{\includegraphics*{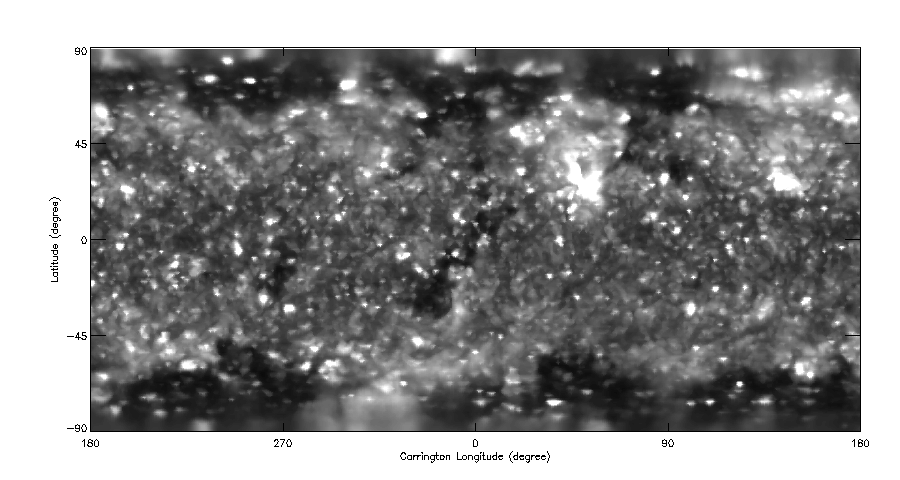}}
\resizebox{0.45\textwidth}{!}{\includegraphics*{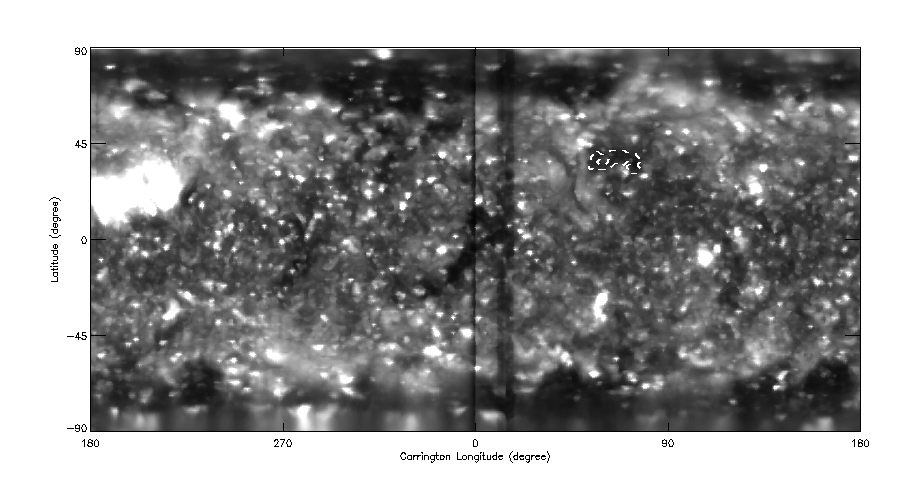}}
\resizebox{0.45\textwidth}{!}{\includegraphics*{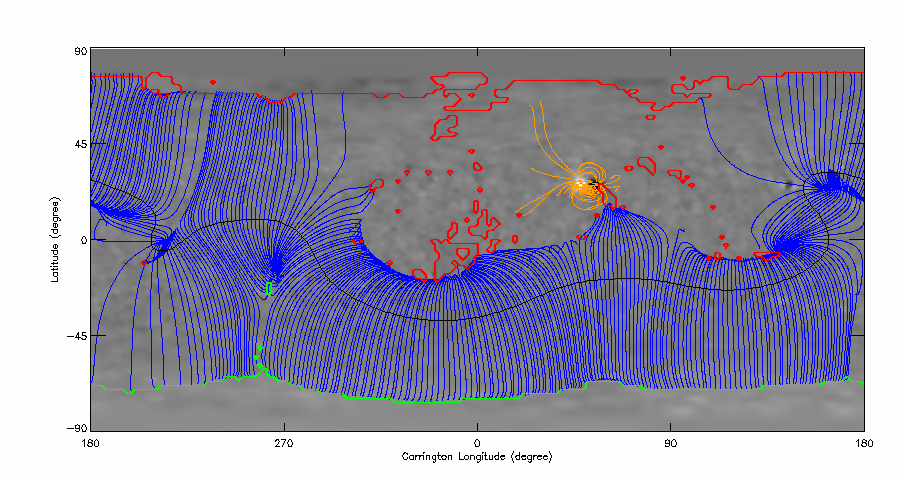}}
\resizebox{0.45\textwidth}{!}{\includegraphics*{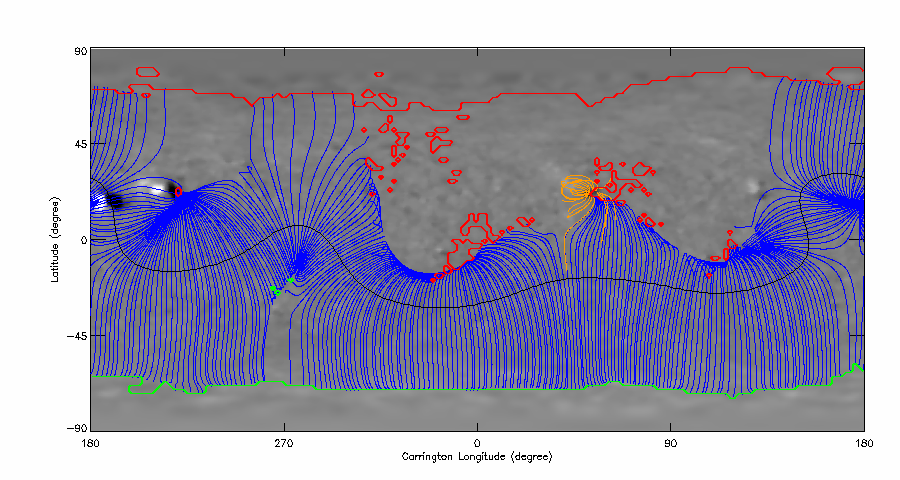}}
\resizebox{0.45\textwidth}{!}{\includegraphics*{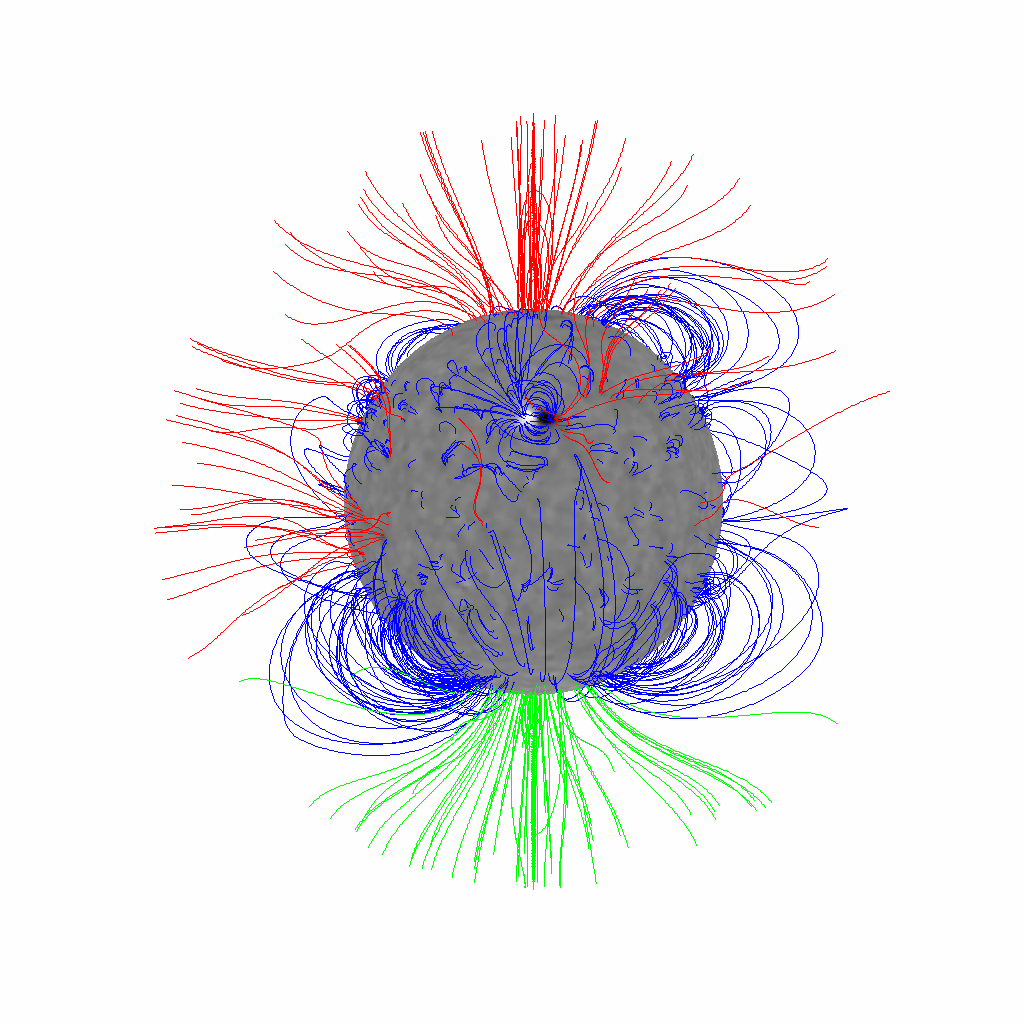}}
\resizebox{0.45\textwidth}{!}{\includegraphics*{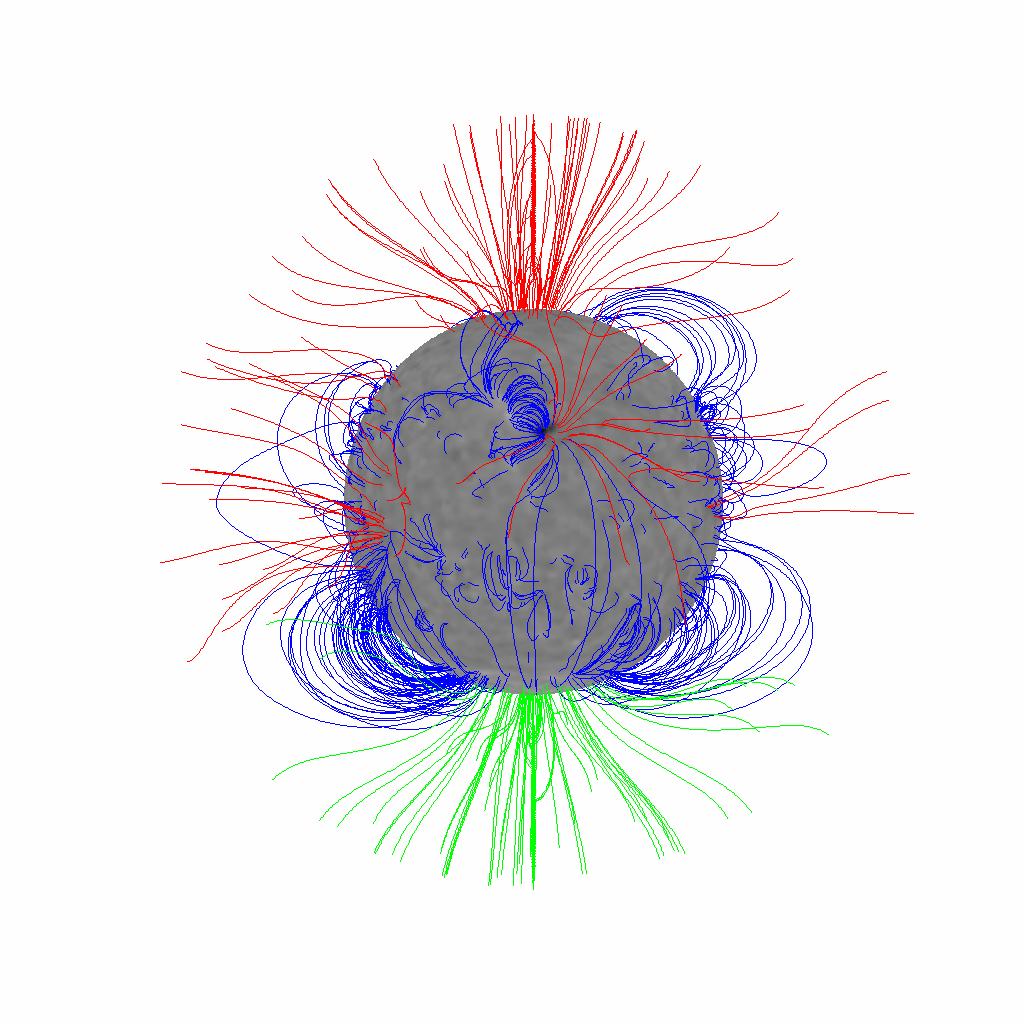}}

\vspace{-1.5cm}

\end{center}
\caption{\footnotesize Example 6 of Table~\ref{arlist}. Top row: GONG synoptic magnetograms for Carrington rotations 2081 and 2082 with positive/negative flux in bright/dark greyscale. The decaying active region is indicated by the black contour curves. Second row: the STEREO EUVI 195~\AA\ synoptic maps for Carrington rotations 2081 and 2082. The coronal hole is indicated by the white dashed EUV brightness contour curve. Third row: PFSS models over-plotted on the synoptic magnetograms. The green and red contours indicate foot-point locations of positive and negative coronal holes. The blue lines represent the streamer belt, and the black source-surface neutral line the equatorial current sheet. The yellow lines represent the active region fields whose foot-points lie within the contour curves shown in the top row of plots. Plots of the same models in spherical coordinates are shown in the bottom row. Closed lines are plotted in blue and positive/negative open lines in green/red. The spherical coordinate system is rotated so that the decaying region is at central meridian.}
\label{fig:ex6}
\end{figure}


\begin{table}
\begin{center}
\caption{Properties of Active Regions}\
\label{arlist}
\footnotesize
\\
\begin{tabular}{lclllllll} \hline\hline
Ex./ & Carr. & Carr. lon, & $^3$d$x$,d$y$  & Hale/  & Flux ($10^{20}$~Mx), & $\overline{|{\bf B}|}$ (G)  & $^4r_{\rm eff}$ (Mm) & CH, \\
Rot. & Rot.    & lat ($^\circ$)  &  (Mm)      & Joy & pos, neg, \% imb & pos, neg & all, pos, neg & polarity \\ \hline
1/1 &	2051 &	16,-6 &	-4,9 &		Y/Y &		201,-288,-18\% &	19,-27 &		55,77,43 &	Y,- \\
1/2 &	2052 &	23,-7 &	24,33 &		N/N &		166,-244,-19\% &	21,-21 &		59,85,51 &	Y,- \\
1/3 &	2053 &	33,-9 &	106,34 &		N/N &		78,-138,-28\% &	18,-12 &		61,64,62 &	Y,- \\
1/4 &	2054 &	29,-8 &	167,56 &		N/N &		53,-83,-22\% &	15,-9 &		61,23,68 &	Y,- \\
2/1 &	2052 &	224,1 &	88,19 &		Y/N &		91,-104,-6\% &	28,-27 &		25,25,18 &	Y,- \\
2/2 &	2053 &	229,2 &	110,45 &		Y/N &		23,-47,-34\% &	7,-10 &		31,34,22 &	Y,- \\
3/1$^{1,2}$ &	2057 &	228,-13 &	-75,-40 &		Y/Y &		36,-39,-4\% &	17,-20 &		17,11,10 &	Y,- \\
3/2$^{1,2}$ &	2058 &	234,-14 &	-135,-118	 &	Y/Y &		4,-14,-57\% &	7,-8 &		24,7,12 &		Y,- \\
4/1 &	2068 &	234,-10 &	-89,-21 &		Y/Y &		99,-85,7\% &	31,-87 &		18,12,5 &		N \\
4/2 &	2069 &	239,-10 &	-99,-21 &		Y/Y &		32,-19,25\% &	11,-17 &		19,14,7 &		Y,+ \\
4/3 &	2070 &	239,-9 &	--,-- &		--/-- &		12,0,100\% &	8,0 &		15,15,-- &		Y,+ \\
5/1$^2$ &	2075 &	116,26 &	-60,28 &		Y/Y &		23,-25,-3\% &	23,-22 &		12,8,6 &		N \\
5/2$^2$ &	2076 &	117,32 &	-36,24 &		Y/Y &		21,-22,-3\% &	25,-14 &		13,6,15 &		N \\
5/3$^2$ &	2077 &	101,35 &	-91,-10 &		Y/N &		7,-11,-21\% &	7,-10 &		17,11,11 &	Y,- \\
5/4$^2$ &	2078 &	88,36 &	--,-- &		--/-- &		0,-2,-100\% &	0,-6 &		6,--,6 &		Y,- \\
6/1 &	2081 &	50,25 &	-79,7 &		Y/Y &		22,-19,6\% &	13,-17 &		14,9,8 &		Y,- \\
6/2 &	2082 &	49,23 &	-121,47 &		Y/Y &		3,-11,-59\% &	6,-11 &		17,7,8 &		Y,- \\
7/1$^{1,2}$ &	2083 &	199,20 &	-81,2 &		Y/Y &		94,-90,2\% &	20,-23 &		45,42,44 &	Y,+ \\
7/2$^{1,2}$ &	2084 &	201,22 &	-198,-1 &		Y/N &		41,-25,24\% &	11,-9 &		49,40,40 &	Y,+ \\
7/3$^{1,2}$ &	2085 &	181,23 &	--,-- &		--/-- &		11,0,100\% &	7,0 &		13,13,-- &		Y,+ \\
8/1 &	2084 &	295,26 &	-62,21 &		Y/Y &		47,-44,3\% &	30,-34 &		12,8,7 &		Y,- \\
8/2 &	2085 &	289,28 &	-130,43 &		Y/Y &		16,-17,-3\% &	11,-10 &		23,10,11 &	Y,- \\
8/3 &	2086 &	290,30 &	--,-- &		--/-- &		0,-2,-100\% &	0,-7 &		5,--,5 &		Y,- \\
9/1$^1$ &	2085 &	247,-26 &	63,35 &		Y/Y &		52,-54,-3\% &	38,-34 &		13,8,8 &		Y,+ \\
9/2$^1$ &	2086 &	242,-27 &	112,57 &		Y/Y &		34,-30,7\% &	14,-13 &		22,13,13 &	Y,+ \\
9/3$^1$ &	2087 &	242,-28 &	53,66 &		Y/Y &		17,0,100\% &	8,0 &		15,15,-- &		Y,+ \\
10/1$^1$ &	2088 &	213,-33 &	86,24 &		Y/Y &		68,-68,0\% &	27,-22 &		18,13,14 &	Y,+ \\
10/2$^1$ &	2089 &	194,-34 &	186,27 &		Y/Y &		27,-31,-7\% &	10,-8 &		32,15,19 &	Y,+ \\
10/3$^1$ &	2090 &	178,-37 &	279,31 &		Y/Y &		8,-5,19\% &	8,-8 &		40,9,7 &		Y,+ \\
11/1 &	2089 &	302,-29 &	75,28 &		Y/Y &		21,-24,-6\% &	22,-15 &		14,6,10 &		N \\
11/2 &	2090 &	301,-31 &	--,-- &		--/-- &		6,0,100\% &		7,0 &		8,8,-- &		Y,+ \\
12/1 &	2090 &	216,18 &	-102,79 &		Y/Y &		83,-82,0\% &	18,-22 &		24,18,15 &	Y,- \\
12/2 &	2091 &	214,22 &	-188,91 &		Y/Y &		41,-42,-1\% &	10,-13 &		36,21,15 &	Y,+,- \\
13/1 &	2099 &	333,19 &	-127,26 &		Y/Y &		158,-160,-1\% &	29,-32 &		28,23,20 &	Y,+ \\
13/2 &	2100 &	338,19 &	-161,18 &		Y/Y &		135,-164,-10\% &	15,-26 &		44,43,29 &	Y,- \\
13/3 &	2101 &	340,21 &	-288,55 &		Y/Y &		57,-101,-28\% &	9,-20 &		54,43,21 &	Y,- \\
14/1 &	2101 &	132,-31 &	147,28 &		Y/Y &		162,-173,-3\% &	69,-35 &		26,9,17 &		Y,- \\
14/2 &	2102 &	115,-34 &	217,54 &		Y/Y &		100,-114,-7\% &	21,-16 &		40,21,27 &	Y,+,- \\
14/3 &	2103 &	96,-34 &	174,33 &		Y/Y &		81,-76,3\% &	21,-12 &		42,27,43 &	Y,+ \\
14/4 &	2104 &	73,-37 &	274,12 &		Y/Y &		36,-25,17\% &	10,-7 &		46,18,32 &	Y,+ \\
\end{tabular}
\end{center}
\footnotesize
\hspace*{40pt}$^{1}$Previously analyzed by Karachik et al.~(2010).\\
\hspace*{40pt}$^{2}$Previously analyzed by Wang et al.~(2010).\\
\hspace*{40pt}$^{3}$Displacements in longitude (d$x$) and latitude (d$y$) between positive and negative flux centroids.\\
\hspace*{40pt}$^{4}$See Equation~(\ref{eq:reff}).
\end{table}

\begin{table}
\begin{center}
\caption{Summary of Qualitative Behavior of the Active Regions}\
\label{chlist}
\\
\begin{tabular}{lcccccccc} \hline\hline
Ex. & Solar & Hemisphere/ & $^1$Dense/ & L or T  & dense  & initial/final & $^2$E-ward or & $^3$CH matches \\
      & Cycle & lead polarity & compact & opens  & opens    &  imbalance  &  P-ward of SB & polar field \\ \hline
1 & 23 & S/- & L/L & L & Y & -/- & E & N \\
2 & 23 & N/+ & L/T & L & N & -/- & S & Y \\
3 & 23 & S/- & L/L & L & Y & -/- & E & N \\
4 & 23 & S/- & L/L & T & N & +/+ & S & Y \\
5 & 24 & N/- & T/L & L & N & -/- & P & Y \\
6 & 24 & N/- & L/L & L & Y & +/- & P & Y \\
7 & 24 & N/- & L/T & T & N & +/+ & E & N \\
8 & 24 & N/- & L/L & L & Y & +/+ & P & Y \\
9 & 24 & N/- & L/= & L & Y & -/+ & P & Y \\
10 & 24 & S/+ & L/L & L & Y & 0/+ & P & Y \\
11 & 24 & S/+ & L/L & L & Y & -/+ & P & Y \\
12 & 24 & N/- & L/L & L \& T & N & 0/0 & S & Y \\
13 & 24 & N/- & L/L & L & Y & -/- & S & Y \\
14 & 24 & S/+ & L/L & L & Y & -/+ & S & Y \\
\end{tabular}
\end{center}
\hspace*{40pt}$^{1}$Was the leading (L) or trailing (T) flux more dense/compact?\\
\hspace*{40pt}$^{2}$Was the region initially located on the pole-ward (P) or equator-ward (E) side of the PFSS streamer belt, or contained within it (S)? Note that none of the cycle 23 examples was initially located pole-ward of the streamer belt, and only one cycle 24 region was initially located equator-ward of the streamer belt.\\
\hspace*{40pt}$^{3}$Did the coronal hole polarity match the polar field in its own hemisphere? Note that the coronal hole polarity matched the polar field in its own hemisphere in every case except those regions initially located equator-ward of the streamer belt. Thus there was no mismatch between a coronal hole magnetic polarity and the polar field polarity on the original active-region's side of the streamer belt.
\end{table}

Active regions are understood to emerge from the solar interior in the form of closed loops with balanced magnetic flux but with asymmetries between the positive and negative magnetic flux distributions. An active region is usually composed of a magnetic bipole where one polarity clearly trails the other in longitude. It is well known that active regions tend to follow a simple pattern whereby the leading magnetic fluxes of the bipoles in the north and south hemispheres have opposite polarities and that these polarities switch during each activity minimum (Hale and Nicholson~1925), known as Hale's polarity rule. For example, the leading flux in the north/south hemisphere generally had positive/negative polarity during cycle 23 and negative/positive polarity during cycle 24. The active regions also tend to be asymmetric in their flux distribution: the leading flux is usually nearer the equator (Hale et al. 1919, a rule known as Joy's law), and is also more dense and compact than the trailing flux. The active regions' asymmetry has consequences for their decay patterns. The trailing flux tends to decay more quickly than the leading flux and be transported pole-ward, often leaving the leading flux isolated. The leading flux may therefore be expected to open up and form a low-latitude coronal hole under some circumstances (Harvey and Sheeley~1979). The asymmetric behavior of the active regions links active and polar fields together in a single magnetic activity cycle (Babcock 1961). During the ascent of an activity cycle the dominant leading polarity of a hemisphere matches the polarity of that hemisphere's polar field, whereas during the declining phase the opposite is true. The fact that the GONG and STEREO data sets include the declining phase of cycle 23 and the ascent of cycle 24 allows us to investigate whether this difference has consequences for low-latitude coronal hole formation.

Table~\ref{arlist} summarizes the quantitative details of the 14 examples of interest in this paper. Each example spans 2-4 rotations. Every active region in the sample followed Hale's polarity rule when first observed and all but one followed Joy's law for active region tilts. Table~\ref{arlist} also lists the magnetic fluxes and flux densities of the regions during each rotation. We estimate the magnetic flux of a region as follows. We find level curves of field strength 5~G and identify by eye the curves that enclose the active region of interest. In Figures~\ref{fig:ex3} and \ref{fig:ex6} (top panels) the selected level curves for examples~3 and 6 are over-plotted in black on the GONG synoptic magnetograms. We then sum by magnetic polarity the flux contained in these curves. Also shown are the average location, flux density and compactness of each region. We follow Karachik et al.~(2010) in calculating the compactness of the active region magnetic field distribution using their expression for the effective radius, $r_{\rm eff}$, where

\begin{equation}
r_{\rm eff}^2 = \sum [ (x-\bar{x})^2 + (y-\bar{y})^2 ]|B_r (x,y)| / \sum |B_r (x,y)| ,
\label{eq:reff}
\end{equation}

\noindent where $\bar{x}$ and $\bar{y}$ are the average pixel coordinates of the field distribution. We calculate $r_{\rm eff}$ for the entire flux distribution of each region and also for the positive and negative fluxes separately.

Table ~\ref{arlist} shows that the total magnetic fluxes of all 14 regions decreased over time, consistent with decay. The average flux densities $\overline{|{\bf B}|}$ generally decreased as the regions decayed but sometimes $\overline{|{\bf B}|}$ increased when some weak flux decayed to values below the threshold used in the analysis and was no longer included in the calculation. The effective radii $r_{\rm eff}$ tended to increase in time as the flux decayed and became more diffuse, until one of the polarities decayed away. The active regions initially had magnetic fluxes ranging from about $4\times 10^{21}$~Mx for a late cycle 23 region to $5\times 10^{22}$~Mx for an early cycle 24 region. The late cycle 23 regions were larger than the early cycle 24 regions on average. The first two regions from cycle 24 were small but the regions subsequently increased in size. The last two regions of the 14 were the second and third largest of the study. The larger regions tended to last longer than the smaller regions. A notable exceptions to this rule was Example~5, which was one of the smallest of the 14 but lasted four rotations.

Figures~\ref{fig:ex3} and \ref{fig:ex6} show GONG and STEREO magnetograms for two typical examples from Table~\ref{arlist}, examples~3 and 6. The black contour curves over-plotted on the GONG maps indicate the decaying active region fields, and the associated coronal holes are indicated by dashed white contour curves on the STEREO synoptic maps. The leading fluxes, negative in both cases, were initially denser than the trailing fluxes and the negative fluxes decayed more slowly, creating local negative flux imbalances in both cases, from which coronal holes developed as shown in the plots. This behavior is much as expected from the previous work cited in the first paragraph of this section. Karachik et al.~(2010) studied example~3 and reported the same behavior. Wang et al.~(2010) analyzed and simulated example~3 and found that the small leading-polarity hole that formed near the equator was balanced by the formation of a trailing-polarity hole near longitude 180, which subsequently developed into an extension of the south polar hole (see the arrowed features in Figure~2 of Wang et al.).  This positive-polarity hole was located in the trailing flux of the active region just to the east of the active region that contained the leading-polarity hole. In this case the two bipoles were close enough together that their facing polarities connected to each other, leaving their outer polarities to open up.  Coronal holes form under the constraint that new open flux of one polarity must always be balanced by opening up an equal amount of flux of the opposite polarity, or by closing down an equal amount of flux of the same polarity. This process can be seen in Figure~\ref{fig:ex3}.

Of the three types of coronal hole identified by Harvey and Recely~(2002), all holes studied here appear to have belonged to the category of holes forming gradually in association with decaying active regions. Some of these regions produced minor flaring activity but, browsing through the STEREO/EUVI sky images, we did not find evidence of a significant eruption associated with sudden formation or development of the holes. The decaying regions appeared to evolve steadily and gradually. In some cases the fully-formed coronal hole would rotate onto the disk where no evidence of coronal hole formation was visible during the previous rotation. In other cases the coronal hole evolved gradually over multiple rotations.

Table~\ref{chlist} summarizes the qualitative behavior of all 14 regions. From this table some clear patterns emerge. In all of the active regions except one, the leading polarity was stronger than the trailing polarity when the region first appeared in a synoptic map. The lone exception, example~5, had marginally more compact leading flux than trailing flux, and it was the leading polarity that was associated with the hole in this case. Example 5 was analyzed and simulated by Wang et al.~(2010), who concluded from their models that the formation of the leading-polarity coronal hole was accompanied by some of the negative-polarity flux from the nearby polar hole closing down by connecting to the trailing flux of the mid-latitude active region via interchange reconnection, as is common during the ascending phase. The cause of the fast decay of positive flux between the times that the CR2076 and CR2077 measurements were taken (see Table~\ref{arlist}) may be related to the distribution of the fluxes according to the GONG synoptic maps. The negative flux was slightly larger but significantly less dense during CR2076, and was composed of two concentrations to the west and south of the positive flux concentration. There were therefore two neutral lines separating the positive and negative fluxes, which may have allowed much flux cancellation to take place during this time interval. The positive flux may have survived the negative flux simply because there was slightly more of it to begin with. The fact that the net flux of the region stays within the range $2-4\times 10^{20}$~Mx throughout four rotations of decay suggests that flux cancellation was the dominant process in this case.

In all but two cases the leading polarity had the more compact distribution. In the exceptional cases, examples~2 and 7, the leading polarity was the more dense on average but, whereas the leading polarity decayed more slowly and opened in example~2, it was the trailing polarity that survived and opened in example~7. Example 7 was also analyzed and simulated by Wang et al.~(2010), who explained the formation and rapid decay of this hole in terms of interchange reconnection between the trailing polarity of the active region and the polar hole boundary. This trailing-polarity (positive) hole (which can be seen in the CR2085 EUVI/A map in Figure~6 of Wang et al. 2010, and which we will model in Section~\ref{sect:pfss}) is associated with the leading negative flux of the neighboring region just eastward that produced a leading-polarity hole. Wang et al.~(2010) showed the roles of diffusion and differential rotation in the formation and evolution of the two holes.

We cannot generally find observations that catch the active regions at their moment of emergence when their flux is expected to be balanced. As soon as a region emerges it interacts with its surroundings and some reconnection will inevitably take place. According to the estimates in Table~\ref{arlist} the earliest observed fluxes were balanced to within 10\% in all regions except one. The exception was the first example which was already a large, mature region at the beginning of the sequence of observations. In most but not all cases the two polarities decayed asymmetrically so that one polarity dominated by the time the coronal hole appeared in the STEREO data. In the two cases where the fluxes remained approximately balanced, examples~10 and 12, the two polarities had become very distant from each other over time, resulting in an isolated unipolar region of field where a coronal hole could form.

In 12 out of 14 cases the leading flux opened, whereas the denser polarity opened in only ten out of 14 cases and the sign of the initial flux imbalance matched the final flux imbalance in only eight out of 14 cases. This indicates a strong hemispheric polarity preference in the decay rates, a pattern that is better correlated with the polarities of the coronal holes than the intrinsic asymmetries of the active regions. On the other hand, the polarity of the coronal hole matched the polarity of the polar field in its hemisphere in only ten out of 14 cases, showing that the hemispheric preference is not strict.

From the ten cycle 24 active regions the leading flux formed the coronal hole in eight cases. The exceptions are examples~7 and 12. In example~12 a coronal hole appeared in the STEREO CR2091 map near the trailing flux. This may seem surprising because in this example the leading flux was initially stronger and more compact. This trailing flux seemed to open following the emergence of two active regions to the east of the decaying region. The emergence of these regions may therefore have played a key role in the formation of the coronal hole from the trailing flux (see Section~\ref{sect:pfss}). In example~7 the leading negative flux was initially denser and more compact than the trailing flux, and yet the leading flux decayed faster creating a positive flux imbalance and a coronal hole. Again the photospheric context of the region will be necessary to explain the formation of this hole (Wang et al.~2010). We will return to both examples~7 and 12 in Section~\ref{sect:pfss}. In each of the remaining eight cases from cycle 24 the leading polarity formed the coronal hole, even when the leading flux was not the denser or more compact polarity.

The data suggest that the global field structure may have a powerful influence over the polarity distribution of coronal holes. Tables~\ref{arlist} and \ref{chlist} show that the coronal holes tend to divide by polarity between the two hemispheres, with negative in the north and positive in the south, matching the polarities of the polar fields. This occurred in most but not all cases: 11 out of 14. During the ascent of an activity cycle the compact and dense polarity of a typical active region following Hale's polarity rule and Joy's law would match the polarity of the polar field in its hemisphere, and the resulting low-latitude holes would therefore match the polar hole in polarity. Tables~\ref{arlist} and \ref{chlist} hint at a link between the global field structure and the polarity of the field that forms the coronal hole.

On the other hand, during the declining phase of cycle~23 this hemispheric bias seems not to have been in strong effect. The leading fluxes of the first four examples were of opposite polarity to the polar fields in their hemispheres. Also these four active regions were relatively large and were situated closer to the equator. One might therefore expect the relationship of these declining-phase coronal holes to the global field to be qualitatively different from their ascending-phase counterparts. Examples~1 and 3 appear to be analogous to most of the cycle~24 examples in that their leading dense fluxes decayed more slowly, creating an imbalance, and coronal holes form in association with them. Examples~2 and 4 appear to have behaved differently. In example~2 the trailing negative flux was initially the more compact and a negative flux imbalance had developed by the next rotation. Yet, the coronal hole that formed was associated with the positive flux, the only case among the 14 whose flux imbalance and coronal hole polarities did not agree. Example~4 also showed unexpected behavior. Here the leading negative flux was initially the denser and more compact and yet it decayed faster, leaving a positive flux imbalance and a coronal hole. These two examples seem difficult to interpret with magnetograms and EUV images alone. In the following section we will revisit them using PFSS models.

While we only have a small number of statistics, it is apparent that there was a larger hemispheric bias in the polarities of the low-latitude holes during the ascent of cycle~24 than during the decline of cycle~23. The relationship between the active regions and the global coronal field structure is fundamentally different during early phases of the solar cycle compared to late phases. During early/late phases an active region typically has relatively high/low latitude and its lagging flux usually opposes/matches the polarity of the photospheric polar field in its hemisphere. We know that during relatively quiet phases of the solar activity cycle it is the low-order multipoles, in particular the dipole and octupole, that dominate the global coronal field structure, and that these components are well correlated with the photospheric polar fields (Petrie~2013). To search for possible links between the the global coronal field structure and the creation of coronal holes, information for the coronal field is needed. To this end, we will present extrapolated potential field models in the following section.

\section{PFSS models}
\label{sect:pfss}



\begin{figure} 
\begin{center}
\resizebox{0.45\textwidth}{!}{\includegraphics*{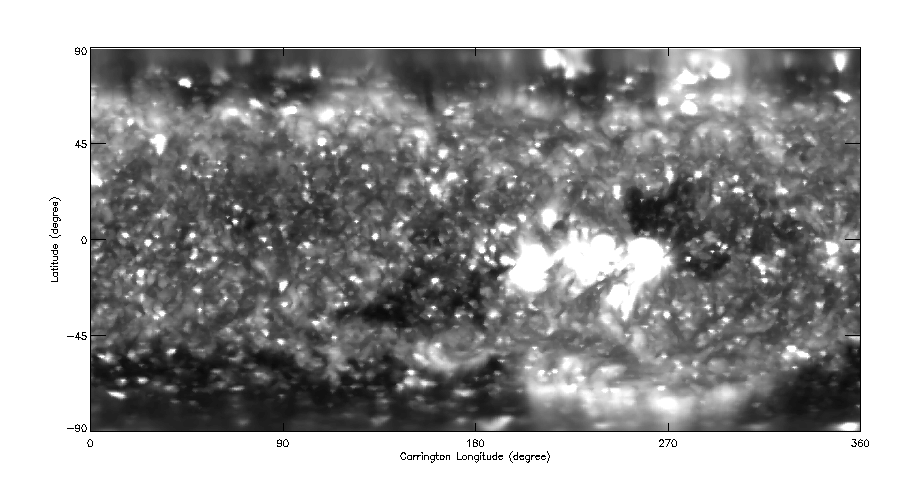}}
\resizebox{0.45\textwidth}{!}{\includegraphics*{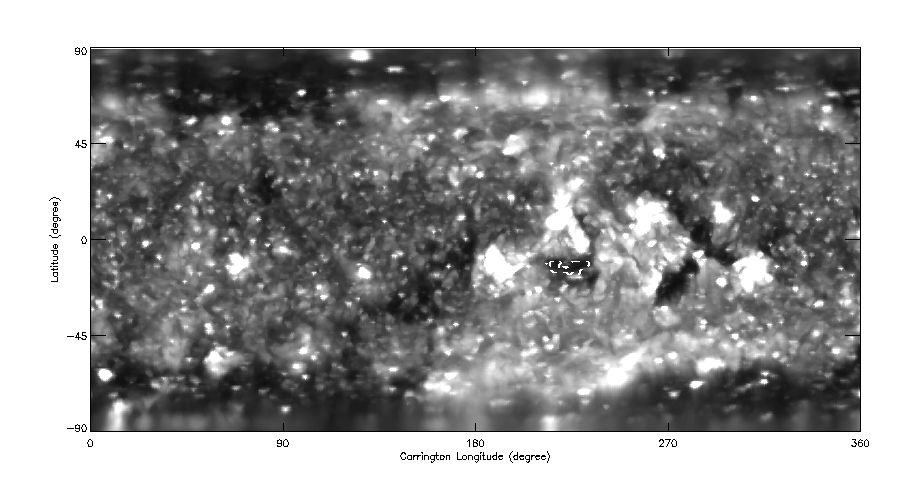}}
\resizebox{0.45\textwidth}{!}{\includegraphics*{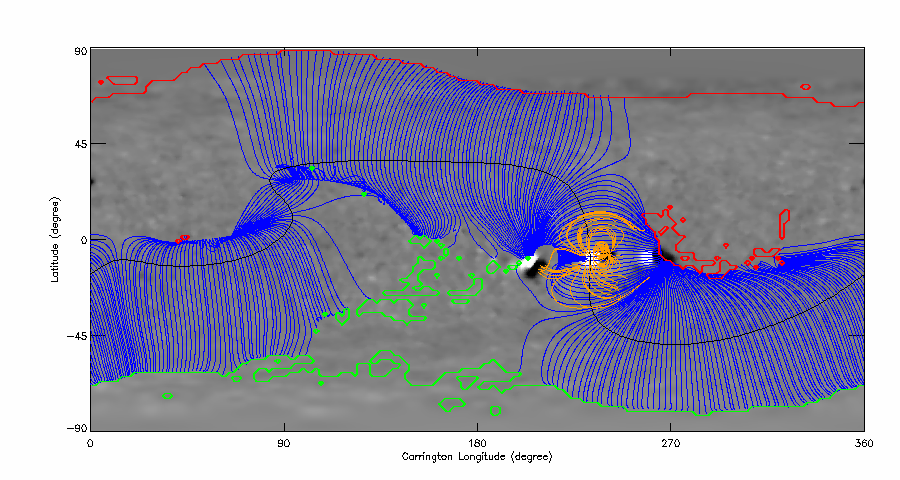}}
\resizebox{0.45\textwidth}{!}{\includegraphics*{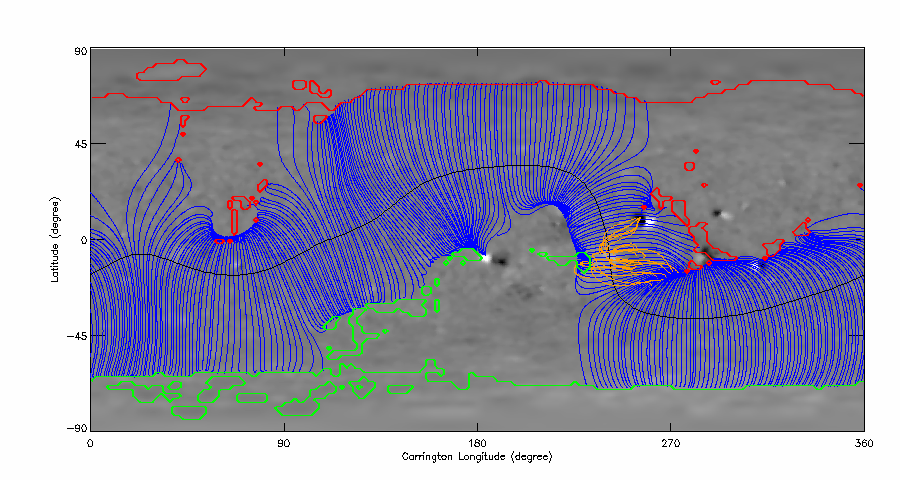}}
\resizebox{0.45\textwidth}{!}{\includegraphics*{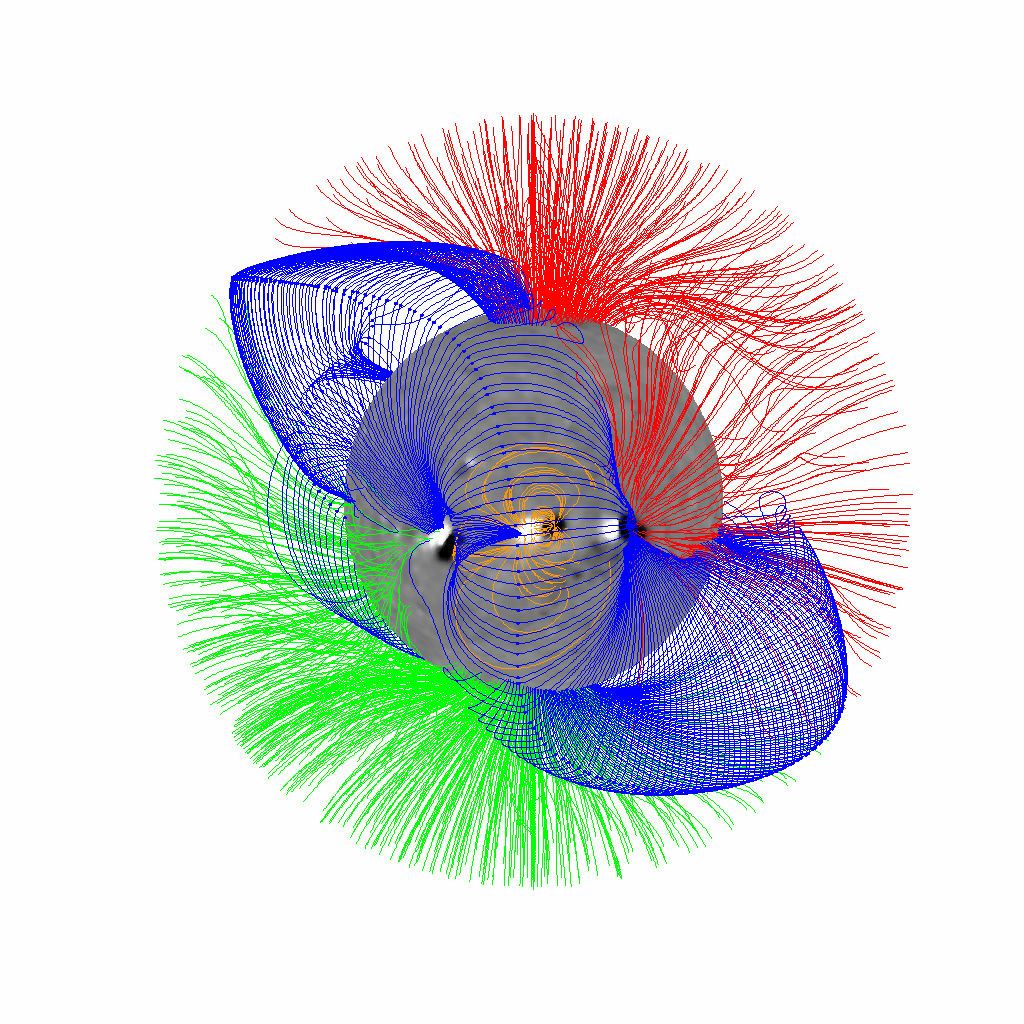}}
\resizebox{0.45\textwidth}{!}{\includegraphics*{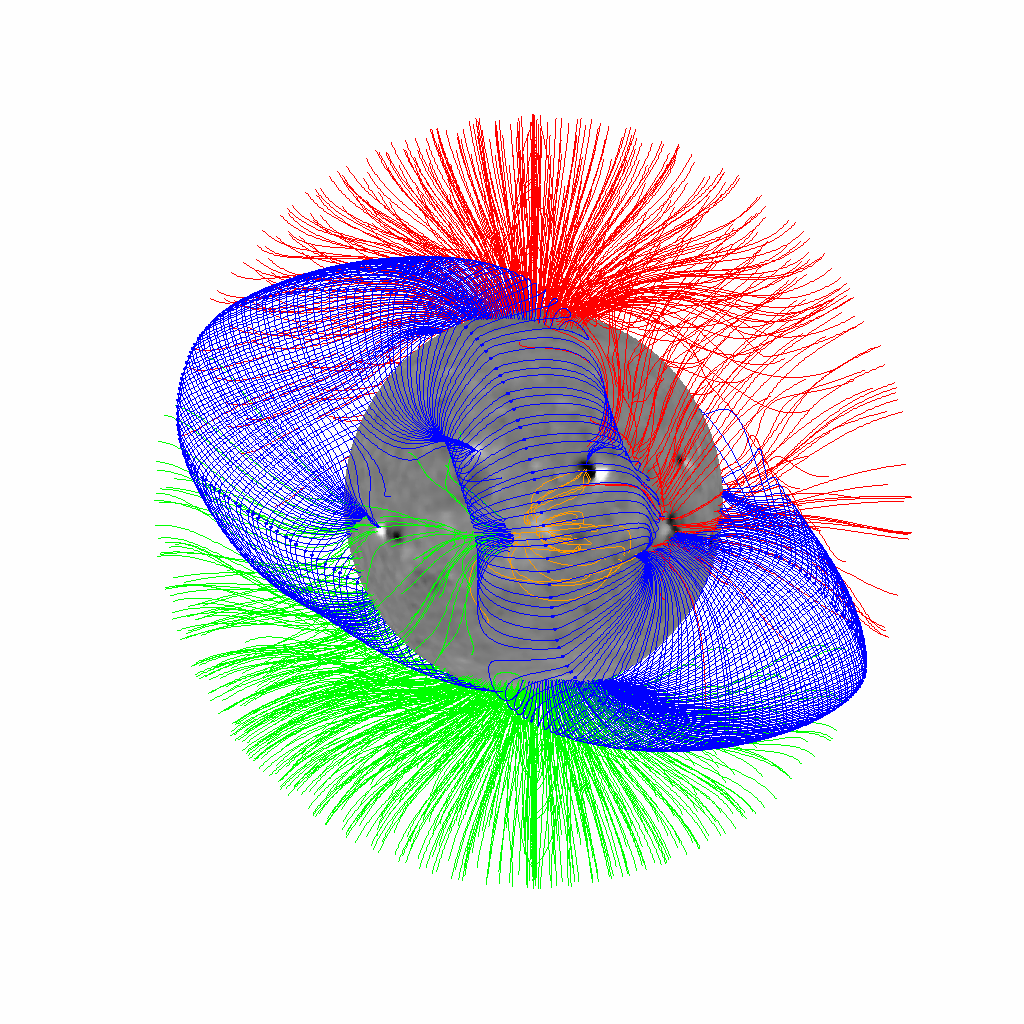}}


\end{center}
\caption{Example 4 of Table~\ref{arlist}. Top row: the STEREO EUVI 195~\AA\ synoptic maps for Carrington rotations 2068 and 2070. The coronal hole is indicated by the white dashed EUV brightness contour curve. Middle row: PFSS models over-plotted on the synoptic magnetograms. The green and red contours indicate foot-point locations of positive and negative coronal holes. The blue lines represent the streamer belt, and the black source-surface neutral line the equatorial current sheet. The yellow lines represent the active region fields whose foot-points lie within the decaying active region. Plots in spherical coordinates of the same features are shown in the bottom row, where the spherical coordinate system is rotated so that the decaying region is at central meridian and open positive/negative field lines are plotted in green/red.}
\label{fig:ex4mod}
\end{figure}

\begin{figure} 
\begin{center}
\resizebox{0.49\textwidth}{!}{\includegraphics*{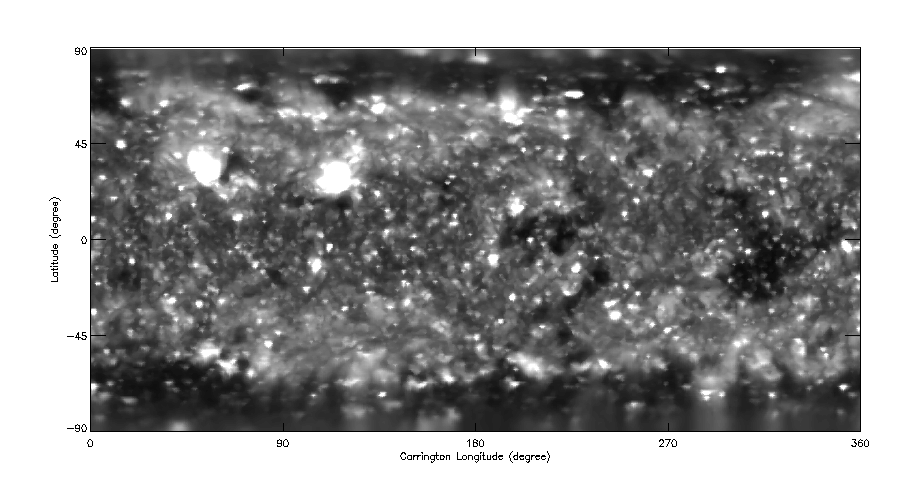}}
\resizebox{0.49\textwidth}{!}{\includegraphics*{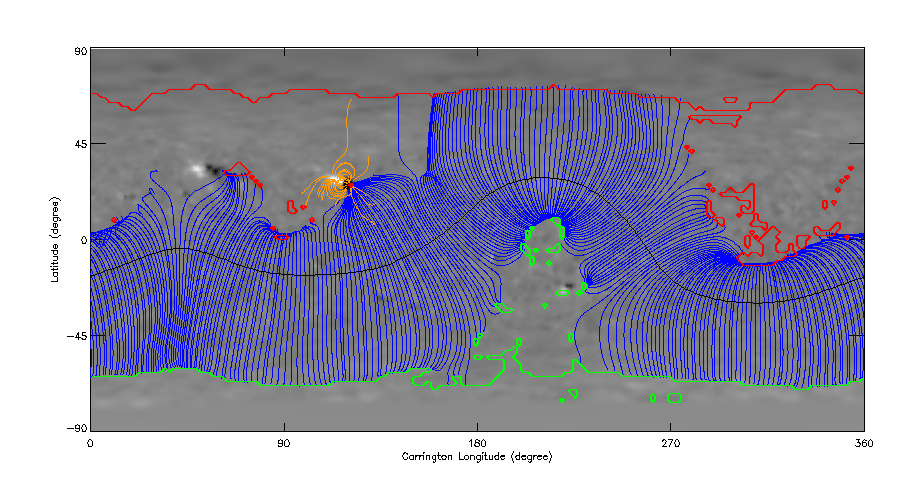}}
\resizebox{0.49\textwidth}{!}{\includegraphics*{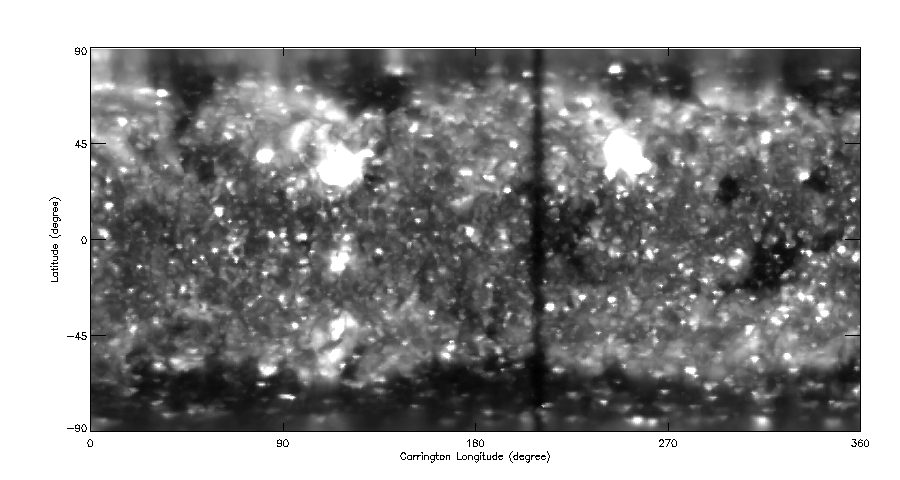}}
\resizebox{0.49\textwidth}{!}{\includegraphics*{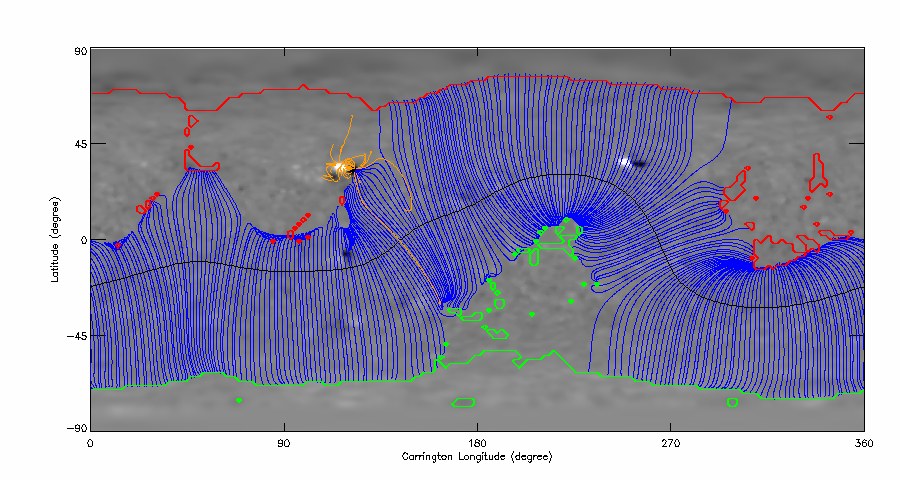}}
\resizebox{0.49\textwidth}{!}{\includegraphics*{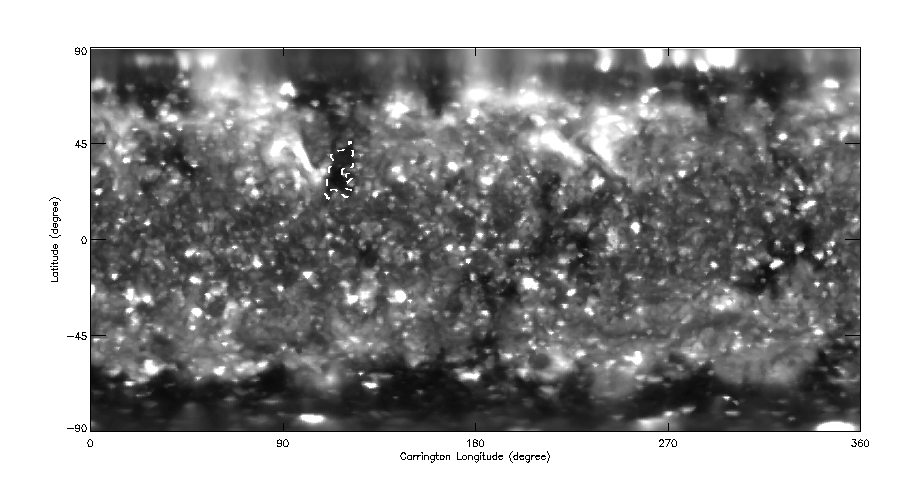}}
\resizebox{0.49\textwidth}{!}{\includegraphics*{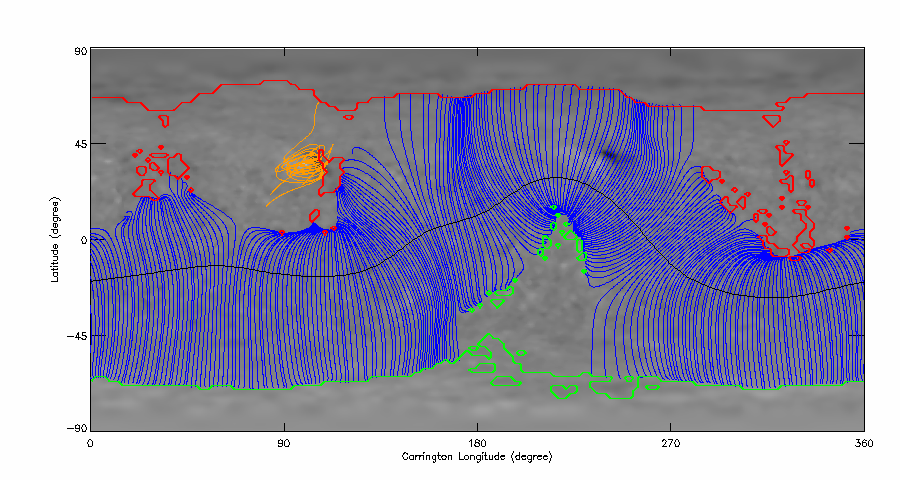}}
\resizebox{0.49\textwidth}{!}{\includegraphics*{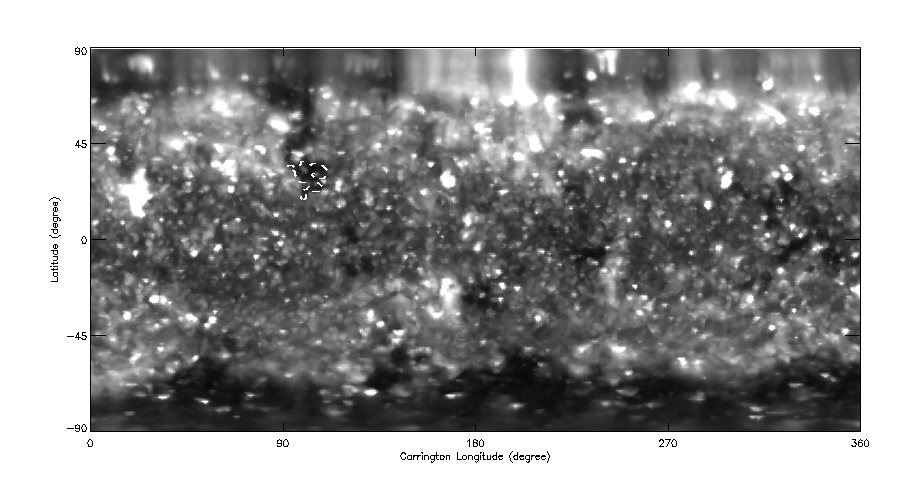}}
\resizebox{0.49\textwidth}{!}{\includegraphics*{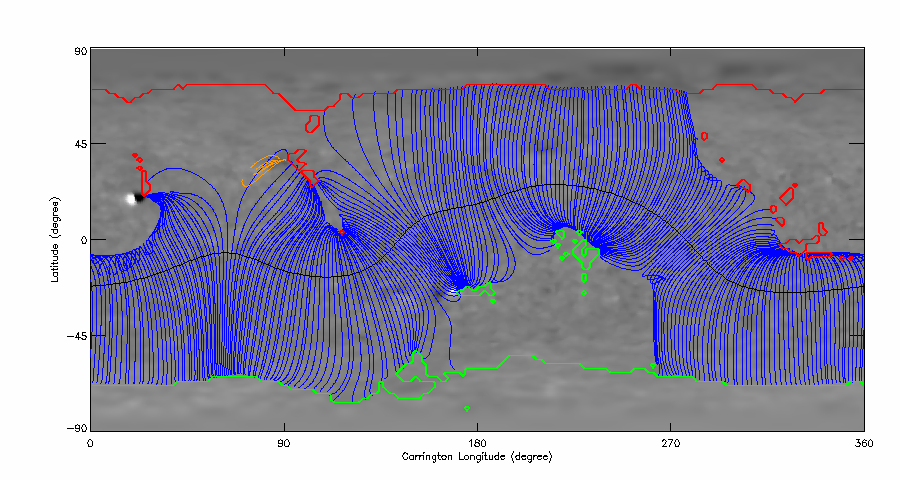}}

\end{center}
\caption{Example 5 of Table~\ref{arlist}. Left column: the STEREO EUVI 195~\AA\ synoptic maps for Carrington rotations 2075-2078. The coronal hole is indicated by the white dashed EUV brightness contour curve. Right column: PFSS models over-plotted on the synoptic magnetograms. The green and red contours indicate foot-point locations of positive and negative coronal holes. The blue lines represent the streamer belt, and the black source-surface neutral line the equatorial current sheet. The yellow lines represent the active region fields whose foot-points lie within the decaying active region.}
\label{fig:ex5mod}
\end{figure}

\begin{figure} 
\begin{center}
\resizebox{0.45\textwidth}{!}{\includegraphics*{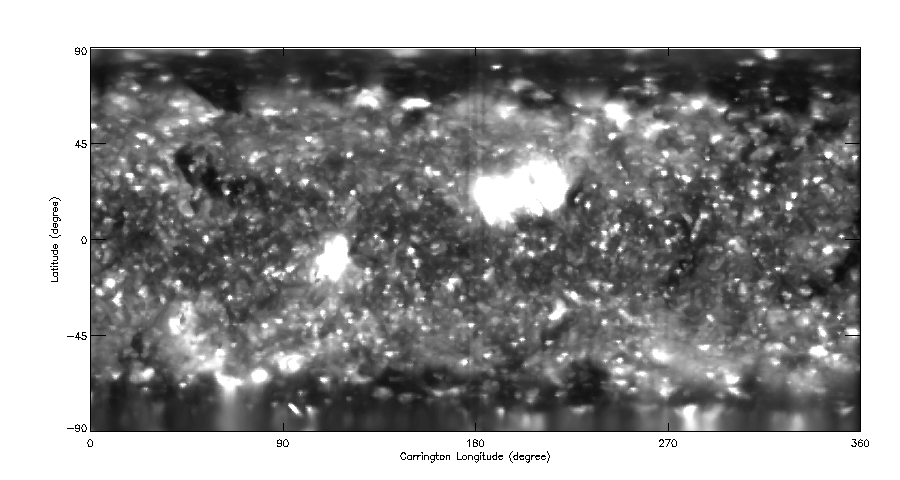}}
\resizebox{0.45\textwidth}{!}{\includegraphics*{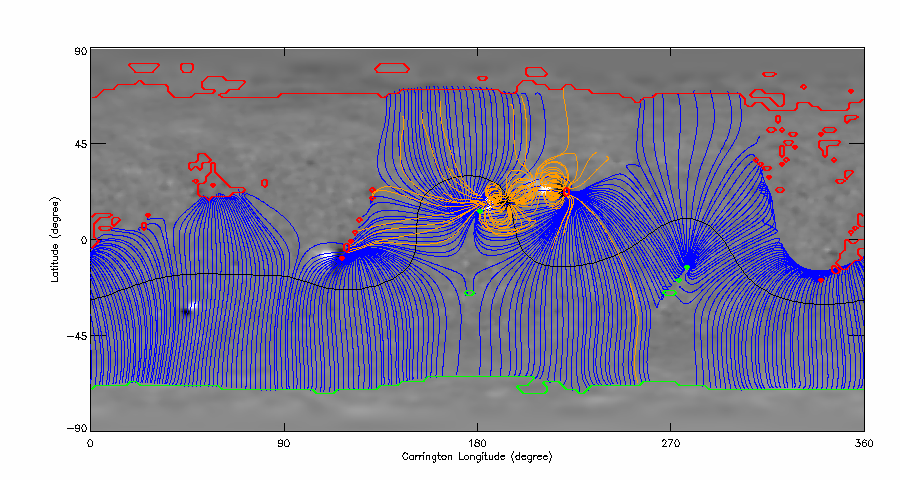}}
\resizebox{0.45\textwidth}{!}{\includegraphics*{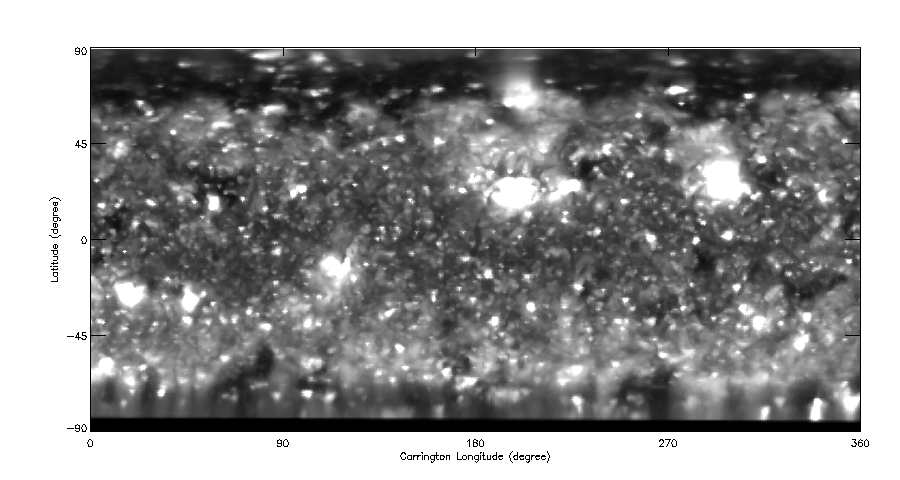}}
\resizebox{0.45\textwidth}{!}{\includegraphics*{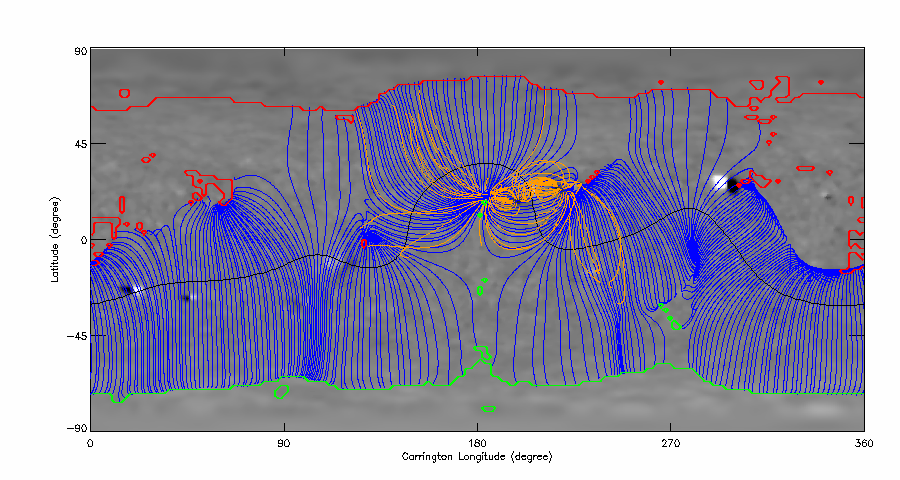}}
\resizebox{0.45\textwidth}{!}{\includegraphics*{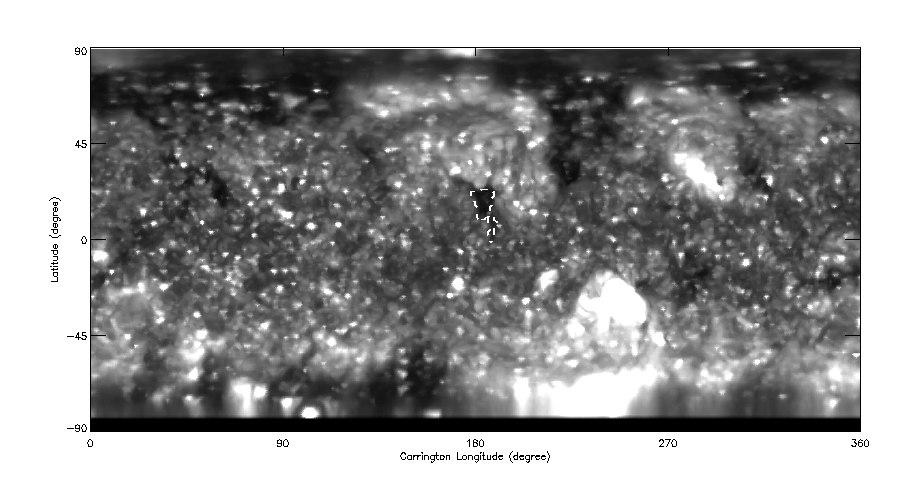}}
\resizebox{0.45\textwidth}{!}{\includegraphics*{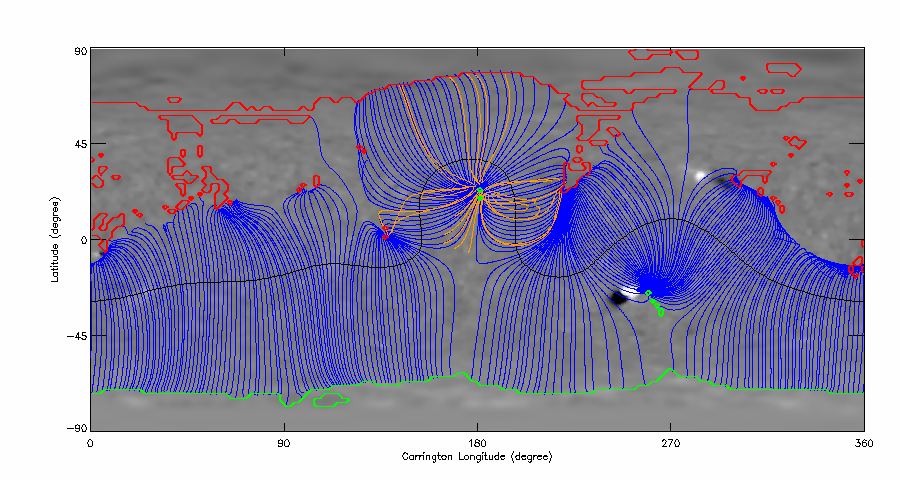}}

\end{center}
\caption{Example 7 of Table~\ref{arlist}. Left column: the STEREO EUVI 195~\AA\ synoptic maps for Carrington rotations 2083-2085. The coronal hole is indicated by the white dashed EUV brightness contour curve. Right column: PFSS models over-plotted on the synoptic magnetograms. The green and red contours indicate foot-point locations of positive and negative coronal holes. The blue lines represent the streamer belt, and the black source-surface neutral line the equatorial current sheet. The yellow lines represent the active region fields whose foot-points lie within the decaying active region.}
\label{fig:ex7mod}
\end{figure}

\begin{figure} 
\begin{center}
\resizebox{0.45\textwidth}{!}{\includegraphics*{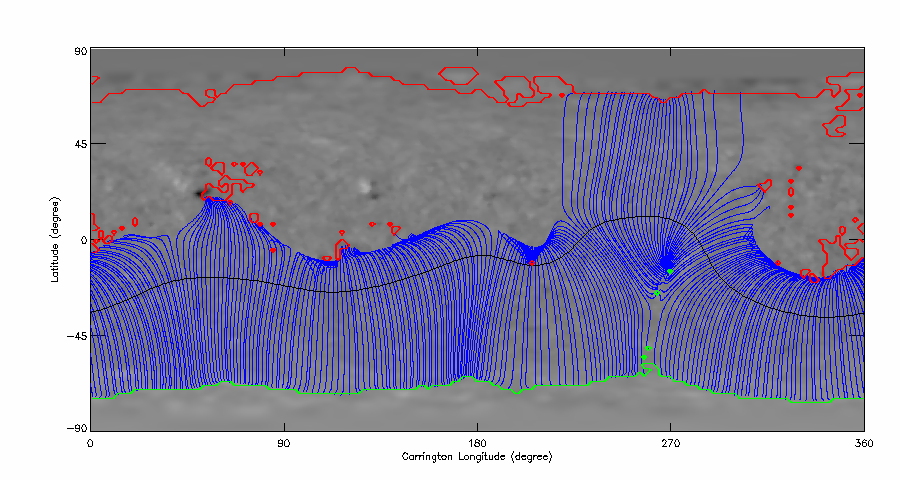}}
\resizebox{0.45\textwidth}{!}{\includegraphics*{fmrnqu090516t1919c2083_000_mask_0_synmag.png}}

\end{center}
\caption{PFSS models over-plotted on the synoptic magnetograms for Carrington rotations 2082 and 2083 (Example~7 of Table~\ref{arlist}). The green and red contours indicate foot-point locations of positive and negative coronal holes. The blue lines represent the streamer belt, and the black source-surface neutral line the equatorial current sheet. The yellow lines represent the active region fields associated with the active region featured in Figure~\ref{fig:ex7mod}.}
\label{fig:ex7emer}
\end{figure}


The EUV images do not tell us the polarity of a coronal hole. Furthermore, as we saw in Section~\ref{sect:arch}, a simple relationship between an EUV dark region and the unbalanced magnetic flux cannot be assumed to determine the polarity of a coronal hole. To supply this and other important information on the coronal magnetic field we extrapolated PFSS models from each GONG synoptic magnetogram. For active regions near the beginnings or ends of Carrington rotations we used composite synoptic maps formed from half of the rotation containing the region and half of the previous or subsequent rotation.

Extrapolating coronal PFSS field models (Altschuler and Newkirk~1969, Schatten et al.~1969) from the GONG synoptic maps is a simple and effective way to diagnose the response of the coronal magnetic field to the photospheric activity patterns described in Section~\ref{sect:arch}. Low in the corona, the magnetic field is sufficiently dominant over the plasma forces that a force-free field approximation is generally applicable. Moreover, for large-scale coronal structure the effects of force-free electric currents, which are inversely proportional to length scale, may be neglected. We identify the lower boundary of the model with the solar photosphere at $r=R$ where $R$ is the solar radius. We use the photospheric radial field maps from GONG to supply the lower boundary data for the radial field component. This use of radial field data, derived from longitudinal measurements of non-force-free photospheric fields, assuming the magnetic vector to be approximately radial in general, has been found to produce models that more successfully reconstruct coronal magnetic structure than the direct application of the longitudinal measurements as boundary data (Wang and Sheeley~1992). A synoptic map construction method that derives the radial component from longitudinal measurements without assuming the fields to be approximately radial has been developed and applied to chromospheric data by Jin et al.~(2013) with promising results. Here we adopt the standard photospheric synoptic maps because measured photospheric fields are found to be approximately radial in general (Svalgaard et al.~1978, Petrie and Patrikeeva~2009, Gosain and Pevtsov~2012), and the resulting models remain useful and competitive for reasons explained by Wang and Sheeley~(1992).

Above some height in the corona, usually estimated to be between 1.5 and 3.5 solar radii, the magnetic field is dominated by the thermal pressure and inertial force of the expanding solar wind. To mimic the effects of the solar wind expansion on the field, we introduce an upper boundary at $r=R_s > R$, and force the field to be radial on this boundary by setting the potential to zero there, following Altschuler and Newkirk~(1969), Schatten et al.~(1969) and many subsequent studies. All fields reaching $r=R_s$ are interpreted as being open to the heliosphere and as representing coronal holes. The usual value for $R_s$ is $R_s=2.5R$ although different choices of $R_s$ lead to more successful reconstructions of coronal structure during different phases of the solar cycle (Lee et al.~2011). In this paper we adopt the value $R_s=2R$  because this value leads to a reasonable visual resemblance between open fields in the models and coronal hole distributions as seen in the STEREO EUV synoptic maps. With these two boundary conditions, the potential field model can be fully determined in the domain $R\le r\le R_s$. We emphasize that each PFSS model is therefore completely determined by the corresponding photospheric magnetogram. There is no time-dependence in the PFSS model. The model is successful because, to a good approximation, the global coronal field relaxes completely to a near-current-free state independent of the previous magnetic field in the sequence.

We computed the PFSS models using the National Center for Atmospheric Research's (NCAR) MUDPACK\footnote{http://www2.cisl.ucar.edu/resources/legacy/mudpack} package. Although PFSS models can be calculated analytically, and have been so treated for several decades, we adopt a finite-difference approach in this paper to avoid some problems associated with the usual approach based on truncated spherical harmonic series (T\'{o}th et al.~2011). MUDPACK (Adams~1989) is a package for efficiently solving linear elliptic Partial Differential Equations (PDEs), both separable and non-separable, using multigrid iteration with subgrid refinement procedures. Jiang and Feng~(2012) have recently presented a high-speed combined spectral/finite-difference PFSS solution method using the related NCAR FISHPACK package for separable PDEs. In this paper we use the MUDPACK package to solve Laplace's equation in its non-separable form in spherical coordinates, subject to the boundary conditions described above.

A successful PFSS model tells us not only the polarity of a coronal hole but also the context of the hole within the global coronal field structure, such as its relationship to the polar coronal holes and the streamer belt. Figures~\ref{fig:ex3} and \ref{fig:ex6} (third rows of plots) show the PFSS models for examples~3 and 6. Here selected fields and features of the model are over-plotted on the GONG synoptic magnetograms. The green and red contours indicate foot-point locations of positive and negative open fields, corresponding to coronal holes. The blue lines delineate the tallest closed field trajectories in the solution, corresponding to the streamer belt, and the black line shows the set of apex locations of these fields, representing the equatorial current sheet. We will refer to this black line as the source-surface neutral line (SSNL) in the following. The yellow lines represent the active region fields whose foot-points lie within the black contour curves of Figures~\ref{fig:ex3} and \ref{fig:ex6} (top plots). The figure also shows plots of the models in spherical coordinates. These plots show magnetic lines selected according to the field intensity at their foot-points. Closed lines are plotted in blue and positive/negative open lines in green/red. The coordinate system is rotated so that the decaying region is at central meridian. The plots show the development of open field at the location of the decaying region and enhanced connectivity to the southern polar latitudes.

The SSNL is generally confined to low latitudes around solar minimum, but major active regions can cause large excursions of the SSNL from the equator. During the decline of cycle 23 and the ascent of cycle 24, the axisymmetric dipole and octupole components of the coronal field had the strongest influence over the global coronal field structure, and these components are well correlated with the slowly-evolving photospheric polar fields (Petrie~2013). During this time, the streamer belt separated the corona into two distinct open-field regions: the global corona had negative/positive open field north/south of the SSNL. It is striking that none of the cycle~23 coronal holes formed pole-ward of the SSNL and only one of the cycle~24 holes, example~7, formed equator-ward, as Table~\ref{chlist} shows. Figures~\ref{fig:ex3} and \ref{fig:ex6} are illustrative of this pattern, showing the equator-ward example~3 from late in cycle 23 and the pole-ward example~6 from early in cycle~24. This is perhaps not surprising behavior for the declining/ascending phase of a solar cycle, when active regions emerge at relatively low/high latitudes. The formation of a coronal hole did not result in a significant change in the location of the SSNL in any of the examples. The low-order magnetic multipole fields, corresponding to large spatial scales, were not significantly perturbed by the photospheric active-region decay associated with the formation of the coronal holes. In each case the hole appeared with the same polarity as the pre-existing open field on its side of the streamer belt. This includes the exceptional case, example~7, where the SSNL looped around the region in the northern hemisphere, placing the region in the southern half of the corona in magnetic terms. In a few late cases, examples~12-14, when the Sun was slightly more active and the regions were more numerous and slightly larger, the corona had a more complex structure and the SSNL overlay the active regions. Under these conditions the decay rates of the positive and negative polarities seem to have been determined by their relative density and compactness.

Examples~1-4 occurred during the declining phase of cycle 23 and were all located near the equator. Two of these holes (examples 1 and 3) appeared equator-ward of the SSNL and two of them (examples 2 and 4) were straddled by it. That none of them occurred pole-ward of the SSNL signifies an important difference between the cycle 23 and cycle 24 coronal holes. Table~\ref{chlist} shows that of the ten out of 14 cases where the polarity of the coronal hole matched the sign of the polar field in its hemisphere, six out of six matched for pole-ward cases and zero out of three for equator-ward cases. These statistics point to a strong influence of the global field structure on the polarity of successfully formed coronal holes. During cycle~23 the leading flux did not match the polarity of the polar field in its own hemisphere, making it much more difficult for it to open and form coronal holes pole-ward of the SSNL than equator-ward. Of the equator-ward cases, in the cycle~23 examples~1 and 3, the leading flux had the same polarity as the polar field on the opposite hemisphere, i.e., the same polarity as the polar field on its side of the streamer belt. These holes were therefore topologically equivalent to cycle 24 holes pole-ward of the streamer belt. A visual comparison of example~3 in Figure~\ref{fig:ex3} and example~6 in Figure~\ref{fig:ex6} illustrates this topological relationship. The major difference between these two examples is that example~3 occurred south of the equator. In terms of magnetic topology they are equivalent. In common with the cycle 24 holes, the equator-ward cycle 23 holes in examples~1 and 3 formed from flux of the same polarity as the pre-existing open field on their side of the streamer belt. In these two cases the leading flux was also more dense and more compact than the trailing flux. The global field structure and the magnetic asymmetry of the active region together seem to have enabled the leading-polarity coronal hole to form.

Of the five cases occurring near the SSNL, the denser polarity formed the coronal hole in only three of them: zero out of two during cycle 23 and three out of three during cycle 24. The three cycle 24 cases appear to have evolved straightforwardly according to their asymmetric flux density distributions. The denser leading polarity survived to form the hole in all three cases. As noted in the previous section, the two examples from cycle 23, examples~2 and 4, are more difficult to explain. Aided by the PFSS models, we do so now.

Example~2 was centered a degree or two north of the equator and had the leading positive polarity of a northern-hemisphere cycle 23 region. The SSNL straddled the active region and the trailing negative flux was initially more compact than the positive flux. A significant negative flux imbalance developed as the positive flux decayed more quickly than the negative flux. According to the PFSS models this flux imbalance manifested mostly as increased magnetic connection to the positive south polar latitudes. The coronal hole that did form near the region had positive polarity even though the flux imbalance of the decaying region was clearly negative (-34\%). A newly emerged region to the west visible in the CR2053 map (not shown here) seems to have been involved in the formation of this coronal hole. Its positive trailing polarity combined with the leading polarity of the decaying region to create an area of positive flux where a coronal hole could develop. This is the only example in our sample of 14 where the polarities of the decaying active region flux imbalance and the associated coronal hole did not match.

As in example~2, in example~4 (Figure~\ref{fig:ex4mod}) the SSNL straddled the active region but this southern region had leading negative flux that was initially much denser and more compact than its positive flux. And yet the flux balance became overwhelmingly positive over time and a coronal hole of positive polarity formed. Why? In this case the positive flux imbalance and associated asymmetric topology seem to explain the asymmetric development of this region. The active region was initially tucked underneath the streamer belt. The streamer belt separated the closed flux underneath from the negative open flux to the north and the positive open flux to the south. Based on its positive flux imbalance, this region might be expected to have had more magnetic connection to the north (negative) side half of the corona than the south, consistent with its positive flux imbalance. According to the model, most of these connections occurred within the streamer belt via lines arching from the positive flux of the region over the western half of the region to the distant negative flux of the neighboring region to the west. These connections were discussed in detail by Petrie et al.~(2009) and are evident in Figure~\ref{fig:ex4mod}. Since the connecting fields extended high in the streamer belt they were particularly susceptible to being reconnected at the SSNL and opened up. According to the plots for the subsequent rotations shown in Figure~\ref{fig:ex4mod}, this seems to be what happened. By the time of CR2070, some of the positive flux of the region was open and a dark coronal hole corresponding to the positive flux was apparent in the EUV map, but the basic global structure of the coronal field was unchanged.

Figure~\ref{fig:ex5mod} shows the STEREO EUV maps and PFSS models for example~5. As noted in the last section, this small region survived several rotations and underwent fast decay between CRs~2076 and 2077, perhaps due to flux cancellation. A sizable coronal hole developed to the west of its main leading-polarity (negative) flux concentration between CRs~2076 and 2077. This coronal hole stands out because it developed in association with negative active region flux that was less dense than the corresponding positive flux in the region and yet decayed more slowly and survived to create the coronal hole. We discussed in the last section that this may have been due to the distribution of the negative flux in two concentrations, bordering the positive flux at two neutral lines, visible in Figure~\ref{fig:ex5mod} allowing flux cancellation to dominate the evolution during CRs~2076 and 2077. Given that there was slightly more negative than positive flux observed in this region from the start of the sequence of GONG observations, flux cancellation alone would ultimately produce a significant negative flux imbalance. The PFSS models show that the resulting negative coronal hole also preserved the global coronal structure as a positive hole in this region could not have done. We found no example where the formation of a coronal hole significantly changed the coronal structure.

The PFSS model plots for example~7 in Figure~\ref{fig:ex7mod} demonstrate the influence active regions can have on the global structure of the corona. The SSNL followed a tight loop-shaped trajectory to straddle the active region whose trailing positive polarity ultimately produced the coronal hole. Throughout the three rotations the SSNL maintained this shape and separated this open trailing positive flux from the open leading negative flux of the neighboring region to the east. The SSNL appears to have been locked in this position as the active regions decayed and the EUV coronal hole formed from this trailing flux. This is in marked contrast to the effect of this region when it emerged into the corona. As Figure~\ref{fig:ex7emer} shows, there was a major change between CR2082, when the SSNL lay south of the equator around $180^{\circ}$ longitude before the region emerged, and CR2083, when the SSNL encroached more than $30^{\circ}$ northward into the northern hemisphere in response to the emergence of the region and its neighbor to the east. The contrasting responses of the global field structure to active region emergence and decay are due to the effects of these processes on the photospheric field distribution and their timescales. When active regions emerge into the atmosphere at a location where no pre-existing active-region flux is present. the effect on the global field can be significant. The low-order multipole moments of the active region can be large enough to perturb the streamer belt over timescales much less than a rotation as in Figure~\ref{fig:ex7emer}. The effects of the decay of this region was much slower: between CRs~2083 and 2085 the region decayed almost completely yet the streamer belt structure hardly changed. The decay processes left a large-scale patch of weak positive flux which, with the large-scale patch of weak negative flux left by the decay of the neighboring region to the east, maintained a low-order magnetic multipole moment in CR 2085 comparable to that of the newly emerged region in CR 2083. This preserved the kink in the streamer belt even after the active-region-strength fields had almost completely decayed away. In this case the streamer belt did not lose its kink until CR2087, by which time the weak unipolar photospheric fluxes of the two regions had been transported from the area. In each of our examples the remnant flux of the decayed active region generally maintained the influence of the region on the global coronal field at least a rotation after the region had decayed.

In several of the examples, in particular 1, 2, 7 and 12, we see evidence in the models that the active region field was open long before a dark coronal hole was observed in EUV images, where the coronal hole did not become visible until the active region decayed. The plots of the models for example~7, shown in Figure~\ref{fig:ex7mod}, and example~12 indicate that the active regions can contain flux that is open but does not appear as dark regions in the EUV images. Luhmann et al.~(2009) showed evidence of contributions to the solar wind from low-latitude coronal holes during early 2007, including the two active regions from examples~1 and 2. In example~12, according to the models, the leading flux was partially open during both CRs~2090-1 but did not appear in the EUV observations as a dark region. Meanwhile a dark coronal hole appeared in EUV in association with the decaying trailing flux as mentioned in the previous section.



The overall behavior of the decaying active regions may be summarized as follows. The structure of the global coronal field, including the streamer belt, was preserved during the decay of the active regions and the formation of the coronal holes. Thus where the active region was decisively on one side of the streamer belt or the other, the dominant polarity of the open field on that side of the streamer belt invariably matched the polarity of the active-region field that formed the hole. In all cases of isolated active regions within the steamer belt, straddled by the SSNL, the intrinsic magnetic asymmetry of the region became important: the denser, more compact polarity was the one that formed the coronal hole. In cases where the region was within the streamer belt and was not magnetically isolated the coronal hole formed in a complex way particular to the magnetic context of the region, but always without changing the global magnetic structure of the corona. A typical late-phase active region is located closer to the equator and is larger than an ascending phase region, making the structure of the global field somewhat less influential in its topological development. The intrinsic properties of the active region and its complex interactions with neighboring regions therefore appear to be more important factors in determining the behavior of declining-phase regions than of ascending phase regions. Nevertheless the global field structure did not change significantly as a result of the decay of even the largest declining-phase regions studied here. Only the emergence of new active regions seem to have produced significant global change.

\section{Conclusion}
\label{sect:conclusion}

Active region emergence can significantly change the global structure of the corona, reshaping the streamer belt, and sometimes creating multiple streamer structures. Even under the relatively quiet conditions prevailing between 2007 and 2010 when the low-order multipoles dominated the global coronal field, major new active regions produced major changes in the global coronal structure (see Figure~\ref{fig:ex7emer}). Yet we have found that active region decay produces no significant change in the the global structure of the corona over multiple rotations, even for large active regions that develop major flux imbalances. The active photospheric fields change significantly from rotation to rotation, as the figures in this paper show. A Carrington rotation is plenty of time for the coronal field to adjust to a changed photospheric field. Furthermore the PFSS model for a given rotation is computed completely independently of the previous rotation. It is therefore striking that the PFSS coronal field models show almost global structure change due to coronal hole formation over multiple rotations. Two facts seem to explain the continuity of global coronal field structure during coronal hole formation.

First, the global coronal field varied gradually over the period of time studied. During the years 2007-2010 the low-order multipoles, in particular the dipole and octupole, had the strongest influence over the global coronal field structure, and these components are well correlated with the photospheric polar fields (Petrie~2013). The stability of the large-scale component of the global coronal field is due mostly to their relationship to the slowly-evolving polar fields. The resulting large-scale dipolar and octupolar coronal fields were generally more globally influential than the higher-order coronal fields associated with decaying active regions. Photospheric flux transport processes produce unipolar bodies of flux of both polarities in both hemispheres, but the coronal fields of decaying active-regions did not have large enough low-order multipole moments or global influence to cancel the dominant multipoles. They could therefore only open under the condition that they did not significantly change the basic coronal structure dominated by these lower-order fields. Unipolar fields of the ``wrong'' polarity generally remained closed (and unstudied). Thus the basic global coronal field structure, dominated by low-order, large-scale, slowly-varying fields, was preserved.

Second, the remnant weak field had a long-lasting effect that remained after each region had decayed away. Decaying regions that produced coronal holes tended to leave large-scale patches of unipolar weak field which took several rotations to disperse after the region itself had decayed away. These remnant fields were much weaker than the original active-region fields but they were spread over a much larger area. The weak remnant fields had low-order multipole moments of comparable strength to those of the original active regions, and thereby approximately preserved the global field signature of the active region long after the region itself has decayed away. This is why the global coronal field structure remained almost unchanged during the process of coronal hole formation.

We can also comment on the role of these decaying regions in the solar activity cycle. In Babcock-Leighton models for the solar cycle (see, e.g., Charbonneau~2010), the global solar field is often idealized as an approximately axisymmetric configuration, with a belt of tilted active regions in each hemisphere and a global coronal SSNL that does not travel far from the equator. In such models the more dense, compact leading active-region fluxes in the two hemispheres interact with each other across the equator while their less dense trailing counterparts decay and are transported pole-ward. Both the leading and trailing fluxes would contribute to the activity cycle in converting active-region toroidal flux into poloidal flux: the leading fluxes would form trans-equatorial poloidal loops across the equator and the trailing flux would join the nearly axisymmetric polar flux. It is perhaps not surprising that the fields studied in this paper do not behave in this simple way. The photospheric field rarely featured matching pairs of active regions in the two hemispheres. When the global corona had a simple SSNL and a nearly axisymmetric appearance the active regions were small and distant from the equator, and they did not interact significantly with fields on the opposite side of the streamer belt. Their unbalanced flux simply opened up, converting nearly-toroidal active-region flux to nearly-poloidal open flux. In large, magnetically isolated region within the streamer belt the less dense and compact polarity of the region tended to decay faster than the other polarity, leaving it unbalanced and with no partner in the opposite hemisphere to connect with. When a major flux imbalance developed in a region some of its flux necessarily connected elsewhere, usually converting nearly-toroidal active region flux into the nearly-poloidal field of the streamer belt or the open fields. In these ways the decaying active regions participated in the solar activity cycle.


\acknowledgements{}

We thank the referee and Alexei Pevtsov for helpful comments. K.J.H. carried out this work through the National Solar Observatory Research Experiences for Undergraduate (REU) site program, which is co-funded by the Department of Defense in partnership with the National Science Foundation REU Program. This work utilizes data obtained by the NSO Integrated Synoptic Program (NISP), managed by the National Solar Observatory, which is operated by AURA, Inc. under a cooperative agreement with the National Science Foundation. The data were acquired by instruments operated by the Big Bear Solar Observatory, High Altitude Observatory, Learmonth Solar Observatory, Udaipur Solar Observatory, Instituto de Astrof\'{\i}sica de Canarias, and Cerro Tololo Interamerican Observatory. The STEREO/SECCHI data used here are produced by an international consortium of the Naval Research Laboratory (USA), Lockheed Martin Solar and Astrophysics Lab (USA), NASA Goddard Space Flight Center (USA) Rutherford Appleton Laboratory (UK), University of Birmingham (UK), Max-Planck-Institut f\"ur Sonnensystemforschung (Germany), Centre Spatiale de Li\'ege (Belgium), Institut d’Optique Th\'eorique et Applique (France), Institut d’Astrophysique Spatiale (France).

\end{document}